\documentclass[twocolumn]{aastex631}
\usepackage[T1]{fontenc}
\usepackage{upgreek}
\usepackage{graphicx}
\usepackage{soul}
\received{}
\revised{}
\accepted{}

\newcommand{\mdot}{\dot{M}}
\newcommand{\mdotin}{\dot{M}_\mathrm{in}}
\newcommand{\mdotout}{\dot{M}_\mathrm{out}}
\newcommand{\mdotedd}{\dot{M}_{\rm Edd}}

\shorttitle{Energy efficiency in NS ULXs}
\shortauthors{Kayanikhoo et al.}

\begin{document}

\title{Energy flow and radiation efficiency in radiative GRMHD simulations of neutron star ultraluminous X-ray sources}

\correspondingauthor{Fatemeh Kayanikhoo} 
\email{fatima@camk.edu.pl}

\author[0000-0003-2835-3652]{Fatemeh Kayanikhoo}
\affiliation{Nicolaus Copernicus Astronomical Center of the Polish Academy of Sciences, Bartycka 18, 00-716 Warsaw, Poland}
\affiliation{Research Centre for Computational Physics and Data Processing, Institute of Physics, Silesian University
in Opava, Bezru\v{c}ovo n\'am.~13, CZ-746\,01 Opava,
Czech Republic}

\author[0000-0001-9043-8062]{W\l odek Klu\' zniak}
\affiliation{Nicolaus Copernicus Astronomical Center of the Polish Academy of Sciences, Bartycka 18, 00-716 Warsaw, Poland}

\author[0000-0002-9202-8734]{David Abarca}
\affiliation{Nicolaus Copernicus Astronomical Center of the Polish Academy of Sciences, Bartycka 18, 00-716 Warsaw, Poland}

\author[0000-0002-3434-3621]{Miljenko \v{C}emelji\'{c}}
\affiliation{Nicolaus Copernicus Superior School, College of Astronomy and Natural Sciences, Gregorkiewicza 3, 87-100, Toru\'{n}, Poland}
\affiliation{Nicolaus Copernicus Astronomical Center of the Polish Academy of Sciences, Bartycka 18, 00-716 Warsaw, Poland}
\affiliation{Research Centre for Computational Physics and Data Processing, Institute of Physics, Silesian University
in Opava, Bezru\v{c}ovo n\'am.~13, CZ-746\,01 Opava, Czech Republic} 

\begin{abstract}
We investigate numerically the energy flow and radiation efficiency of accreting neutron stars as potential ultraluminous X-ray sources (ULXs). We perform ten simulations {in radiative general relativistic magnetohydrodynamics (GRRMHD)}, exploring six different magnetic dipole strengths ranging from 10 to 100 GigaGauss, along with three accretion rates, 100, 300, and 1000  Eddington luminosity units. Our results show that the energy efficiency in simulations with a strong magnetic dipole of 100 GigaGauss is approximately half that of simulations with a magnetic dipole an order of magnitude weaker. Consequently, radiation efficiency is lower in simulations with stronger magnetic dipoles. We also demonstrate that outflow power increases as the magnetic dipole weakens, resulting in stronger beaming in simulations with weaker magnetic dipoles. As a result of beaming, simulations with magnetic dipole strengths below 30 GigaGauss exhibit apparent luminosities consistent with those observed in ULXs. As for the accretion rates, we find that higher accretion rates lead to more powerful outflows, higher kinetic efficiency, and lower radiation efficiency compared to those of lower accretion rate simulations. 

\end{abstract}


\section{Introduction} \label{sec:intro}

Ultraluminous X-ray sources (ULXs) have been a mystery since their discovery in the 1990s. These objects exhibit luminosities exceeding $10^{39}\,{\rm erg\,s^{-1}}$, surpassing the Eddington limit for stellar-mass compact objects, yet falling well below the typical X-ray luminosities of active galactic nuclei (AGNs). \citet{walton2022} classified a total of 1843 ULXs across 951 host galaxies using data from the \textit{XMM-Newton}, \textit{Swift}, and \textit{Chandra} observatories. Several sources in this catalog display luminosities greater than $10^{41}\,{\rm erg\,s^{-1}}$.

A variety of models have been proposed to explain the physical mechanisms responsible for the luminosity of ULXs. One of the earliest models \citep{Colbert1999} suggests that ULXs are powered by sub-Eddington accretion on intermediate-mass black holes (IMBHs) with masses in the range of $10^2$ to $10^4\,M_{\odot}$. However, this model does not explain the sources with luminosities exceeding $3\times 10^{40}\,{\rm erg\,s^{-1}}$.
\citet{begelman2002} proposed the photon bubble instability model to explain the high luminosities of ULXs powered by stellar-mass black holes with masses $M/M_{\odot} \leq 10$. The instability arises due to interactions between X-ray photons and the gas within the accretion disk. The momentum transfer from photons to gas particles causes localized heating and expansion, leading to the formation of bubble-like structures. As a result, the escaping radiation can exceed the limits predicted by standard accretion disk theory. However, such models can hardly account for luminosities greater than $10^{40}\,{\rm erg\,s^{-1}}$ without invoking strong beaming effects \citep[see][and references therein]{lasota2024problems}.

Based on the fact that in the presence of strong magnetic fields the electron scattering opacity for X-rays is significantly reduced, enabling large amounts of radiation to escape the system \citep{Canuto1971, Elsner1977, Herold1979, paczynski1992}, the 
sources were proposed to be magnetars. 
\cite{Mushtukov2015} pointed out that achieving (true) luminosities exceeding $10^{40}\,{\rm erg\,s^{-1}}$ requires magnetic field strengths above $10^{14}$\,G. However, magnetic fields observed in neutron stars within X-ray binaries are typically below $10^{13}$\,G. 
In this model it is hard to reconcile the neutron star spin-up with the one observed in ULXs, as shown in \cite{kluzniak-lasota2015, Lasota_2023}. 

\cite{King2001} proposed that the extraordinary luminosity of ULXs relates to the apparent luminosity caused by beaming. They suggest that in super-Eddington accreting stellar mass compact objects such as black holes, neutron stars and white dwarfs, the radiatively driven outflow discussed by \citet{Sunyaev1973} causes beaming.
The inferred luminosity $L_{\rm iso}$, which is the luminosity $L$ multiplied by the beaming factor of $1/b$ (with $b\ll1$ the fraction of the solid angle into which the radiation is emitted) could then easily reach values of $10^{39}-10^{41}\, {\rm erg\,s^{-1}}$ \citep{king2009}.

The discovery of coherent pulsations in ULXs \citep{Bachetti2014}, implying\footnote{The only astronomical objects known to have coherent periodicities lower than about a second are neutron stars.} that
the X-ray pulsation period is the spin period of a  neutron star, 
revived interest in the study of ULXs. The spin-up torques of beamed accreting neutron stars \citep{King2001} were examined in the context of ULXs in the KLK model \citep{king2017, king2019, king2020}. Based on a comparison of the model with the observed luminosity, spin, and spin-up/spin-down of pulsating ULXs, the magnetic field of the neutron star within the KLK model is predicted to range from $10^{10}$ to $10^{13}\,$G, with most results clustering between $10^{10}-10^{11}\,$G \citep[see Table 1.1 in][]{lasota2024problems}. 

The first radiative MHD simulations of disk accretion onto neutron stars were conducted by \cite{takahashi2017} with the magnetic field $10^{10}\,$G. While their study yielded several interesting results, the simulations were limited to short durations, preventing the outflows from reaching a steady state and the results from capturing the mechanisms behind beamed radiation. Subsequently, \cite{abarca+21} conducted the long-term simulations with magnetic field $2\times 10^{10}\,$G and showed that outflows caused beaming of radiation and resulted in apparent luminosity consistent with ULXs while their earlier work \citep{abarca+18} showed that the luminosity of a non-magnetized neutron star cannot exceed the Eddington limit even with a super-Eddington accretion rate of 20~$\mdotedd$. 

In \cite{Kayanikhoo2025} we obtained results with dipole strengths $10^{10}$, $5\times 10^{10}$ and $10^{11}\,$G, showing that powerful outflows in these weak magnetic dipole simulations result in strong beaming and apparent luminosity in the ULXs range. \cite{Inoue_2023} showed that the particular source \textit{Swift\,J0243.6+6124} may have a stronger magnetic field and they also performed simulations of ULXs with a quadrupole magnetic field for the neutron star \citep{Inoue_2024}. 

In the follow-up of our previous work, we now examine the impact of magnetic dipole strength and accretion rate on the radiation energy, beaming and apparent luminosity of the accreting magnetized neutron star.

This paper is structured as follows: in Section~\ref{Ssetup} we introduce the
numerical setup.
The magnetospheric radius, accretion, and energy flow are discussed in Section~\ref{Sphysics}. In Section~\ref{Sapparent} are covered the power of outflows and apparent luminosity. In 
Section~\ref{Sconc} we discuss the results and outline the main conclusions.

\section{Numerical setup, initial and boundary conditions}\label{Ssetup}

%
\begin{figure}
    \centering
    \includegraphics[width=\linewidth]{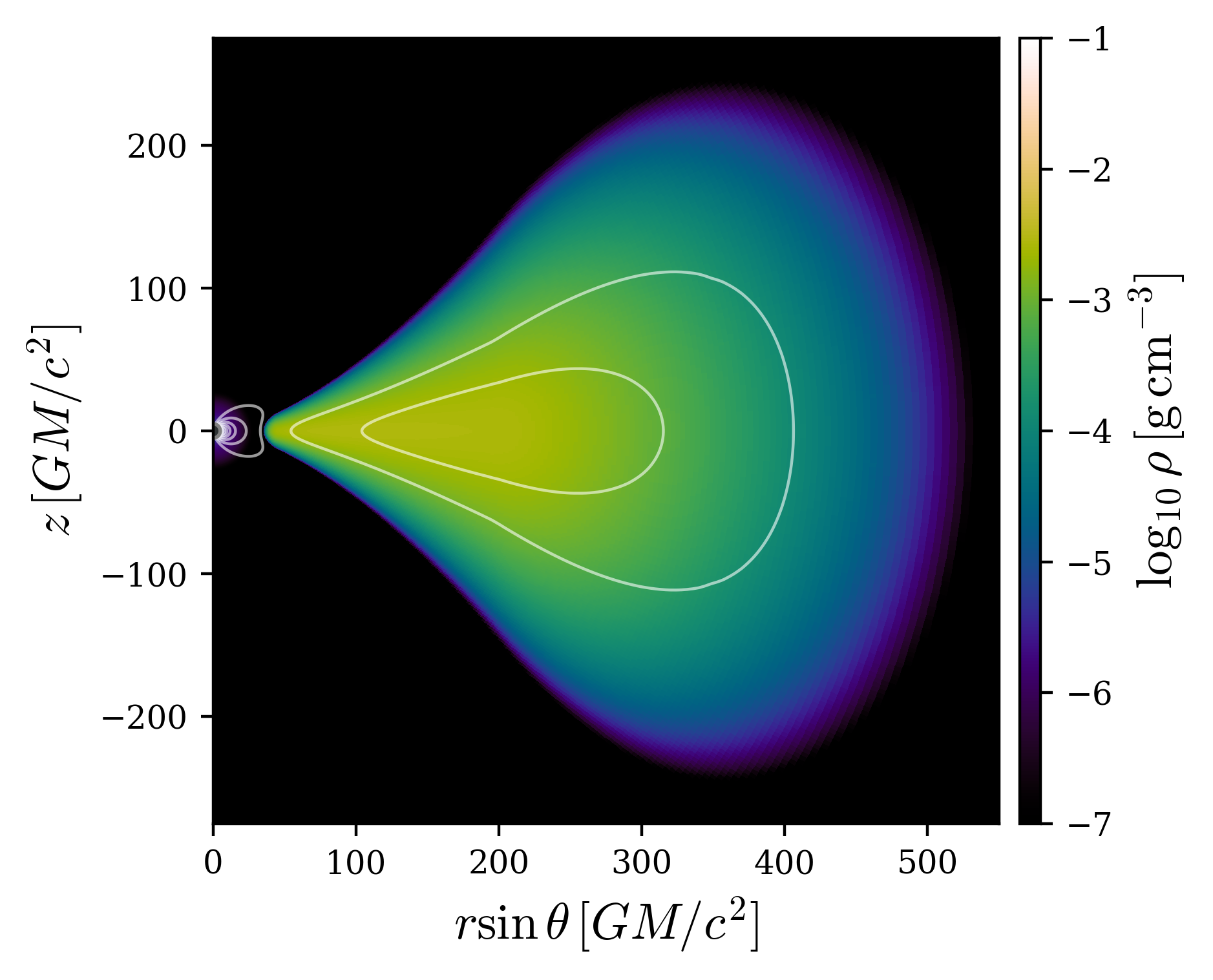}
    \caption{The initial rest-mass density $\rho$ in the fluid frame. Solid grey lines indicate the vector potential $A_\phi$, which in our setup is parallel to the magnetic field lines. The strength of the magnetic field on the surface of the neutron star is $10^{11}$~G.}
    \label{initial_setup}
\end{figure}
Our numerical setup is detailed in \citet{dabarca_thesis, kayanikhoo_thesis} and references therein.
Here we give a brief recapitulation for the reader's convenience. We use the \texttt{Koral} code, which solves the equations of general relativistic radiative magnetohydrodynamics (GRRMHD) on a static mesh with a fixed Schwarzschild metric. 
The equations of conservation of matter, energy-momentum and radiation energy-momentum are solved separately using standard explicit methods for gas and the $M_1$ closure scheme for radiation.
\begin{equation}\label{eqdensity}
    \nabla_\mu (\rho u^\mu) = 0,\quad \nabla_\mu T^\mu\phantom{}_\nu = G_\nu,\quad \nabla _\mu R^\mu\phantom{}_\nu = -G_\nu.
\end{equation}

The gas density in the co-moving fluid frame is denoted by $\rho$, and $u^\mu$ represents the gas 4-velocity. The gas stress-energy tensor $T^\mu\phantom{}_\nu$ and the radiation stress-energy tensor $R^\mu\phantom{}_\nu$ are coupled by the radiation four-force $G_\nu$ \citep{mihalas84}, making use of electron scattering, bremsstrahlung, and synchrotron opacities as well as photon-conserving comptonization \citep{sad_nar_2015}. 

A four-vector related to the magnetic field $b^{\mu}$ is defined in terms of the electromagnetic field tensor $F_{\lambda \kappa}$ and Levi-Civita tensor $\epsilon^{\mu \nu \kappa \lambda}$ \citep{Gammie2003}:  
\begin{equation}\label{mag4vector}
    b^\mu = \frac{1}{2} \epsilon^{\mu \nu \kappa \lambda} u_\nu F_{\lambda \kappa},
\end{equation}
and it is evolved using the induction equation 
\begin{equation}\label{magev}
    \partial _t(\sqrt{-g}B^i) = -\partial_j\left(\sqrt{-g}(b^j u^i - b^i u^j)\right),
\end{equation}
where $B^i$ is the magnetic 3-vector. The $b$ 4-vector components are
\begin{equation}
    b^t = B^i u^\mu g_{i \mu},\quad
    b^i = \frac{B^i + b^t u^i}{u^t},
    \label{4beq}
\end{equation}
where $g_{i \mu}$ is metric. The magnetic fluxes are computed using the flux-CT (Flux Constrained Transport) method to ensure the divergence-free condition \citep{toth+00}. The strong magnetizations in the neutron-star magnetosphere are handled using an innovative flooring scheme introduced by \citet{parfrey+17} and extended for radiative simulations in \citet{abarca+21}.

The neutron star has a mass of 1.4~$M_{\odot}$ and a radius of $R=5\,r_{\rm g}$, where $r_{\rm g} = GM/c^2$ is the gravitational radius. The equilibrium torus (Polish donut) around the star is initialized with the adiabatic equation of state. We use the solution from \citet{penna2013}, incorporating radiation as in \cite{abarca+21}. The (numerical) background atmosphere
is defined with density and internal energy, a few orders of magnitude smaller than that in the initial torus.

This work is divided into two parts: first we examine the impact of the magnetic dipole strengths $[10, 20, 30, 50, 70, 100]\,$GigaGauss on the properties of accreting neutron stars where the mass density of the torus is the same in all simulation with different magnetic dipoles, and initialized to produce a certain accretion rate.
Second, we examine the impact of the accretion rate by varying the initial torus mass density, which results in accretion rates of about three values $[100, 300, 1000]\,L_{\rm Edd}/c^2$, denoted with $\mdot_{100}, \mdot_{300}$, and $\mdot_{1000}$, respectively. We repeat these simulations for two magnetic dipole strengths of 10 and $30\,$GG.

We set the dipole stellar magnetic field and the magnetic field in the torus as a single loop, both defined with vector potential, $A_{\phi}$. We measure the dipole strength using the polar magnetic field. For reference, the dipole moment $\mu$ is related to the polar field strength $B_{\rm pole}$ by the relation
$B_{\rm pole} = 2\mu/R^3$,
where $R$ is the stellar radius\footnote{We use the Lorentz–Heaviside units, with a factor $1/\sqrt{4\pi}$ absorbed into the electromagnetic fields.}. 

The initial setup, including the equilibrium torus and the magnetic field lines, is shown in Fig.~\ref{initial_setup}. In this figure, the maximum strength of the magnetic dipole on the surface of the neutron star is $10^{11}\,$G.

The angular momentum transfer during the simulation is facilitated by the magnetorotational instability (MRI) \citep{Balbus1998}. Operating of the MRI requires a weak magnetic field, so our torus is initialized with the ratio $\beta$ of the total pressure of gas and radiation to the magnetic pressure: $\beta = (p_{\rm gas}+p_{\rm rad})/p_{\rm mag} = 10$. The orientations of the stellar magnetic field and the torus field are opposite. During the simulation, the magnetic field lines reconnect, allowing the gas to accrete onto the surface of the neutron star. 

The boundary conditions are identical to those in \cite{abarca+21}. We briefly outline here the most challenging part, the surface of the neutron star. We use the reflective neutron star's surface with an albedo of 0.75.
The gas slides along magnetic field lines and falls smoothly onto the surface of the neutron star, similar to an absorbing boundary \citep{parfrey+17}. The ingoing internal, radiation and kinetic fluxes are treated to change into the radiation flux with the opposite direction  in two ghost cells above the surface of the neutron star. This converts the in-going energy fluxes to out-going radiation flux. We then multiply the radiation flux by the albedo, which indicates the percentage of the radiation energy flux to be reflected from the surface of the neutron star \citep[for more details see][]{abarca+21}. In our simulations, we adopt an albedo of 0.75, implying that 75\% of the incident energy is reflected as radiation. However, this boundary condition becomes inaccurate in regimes of optically thick accretion. In such cases, radiation is trapped and advected inward with the gas, ultimately being absorbed at the inner boundary before it can be reflected on the surface of the neutron star. In the simulations the effective albedo is about two orders of magnitude lower than intended.

The simulations are performed on a two-dimensional grid, assuming axisymmetry (2.5D approach) with $N_r\times N_\theta\times N_\phi \equiv 510\times512\times1$ cells in the computational domain spanning $[5, 1000]\,r_{\rm g}$ in the radial and $[0, \pi]$ in the polar direction. This resolution is sufficient to resolve the MRI \citep[see][]{Angelos+2024}. Spacing in the radial direction is logarithmic, providing accurate resolution of the field lines right next to the surface of the star. In the polar direction spacing is uniform.

In axisymmetric ideal MHD systems, there is no dynamo mechanism to regenerate the magnetic field \citep[for the anti-dynamo theorem, see][]{cowling+33}. Consequently, the magnetic field decays during simulations, causing the accretion to stop in a short time. However, running simulations in 2.5D is computationally much cheaper than in 3D. To address this issue, \cite{sadowski+dynamo} introduced a mean-field dynamo in \texttt{Koral}, which compensates for the missing dynamo in 2.5D simulations and drives the properties of turbulence towards those characteristic of MRI in 3D simulations. This approach enables the simulations to maintain sustained accretion for significantly longer periods, allowing us to explore the long-term behavior of the system.

The simulations run for a duration up to 50,000~$t_{\rm g}$, where $t_{\rm g} = GM/c^3$ is the gravitational time. The corresponding physical time is approximately 0.35~s, which is approximately one-third of the spin period of the pulsating ULXs; therefore, we neglect the spin of the neutron star in our simulations.

In this paper, the results are based on the time-averaged data in the interval of 15,000 to 50,000~$t_{\rm g}$ for all simulations. 

\section{Magnetospheric radius, accretion and energy flows}\label{Sphysics}
\begin{figure*}
    \centering
    \includegraphics[width =0.75\linewidth]{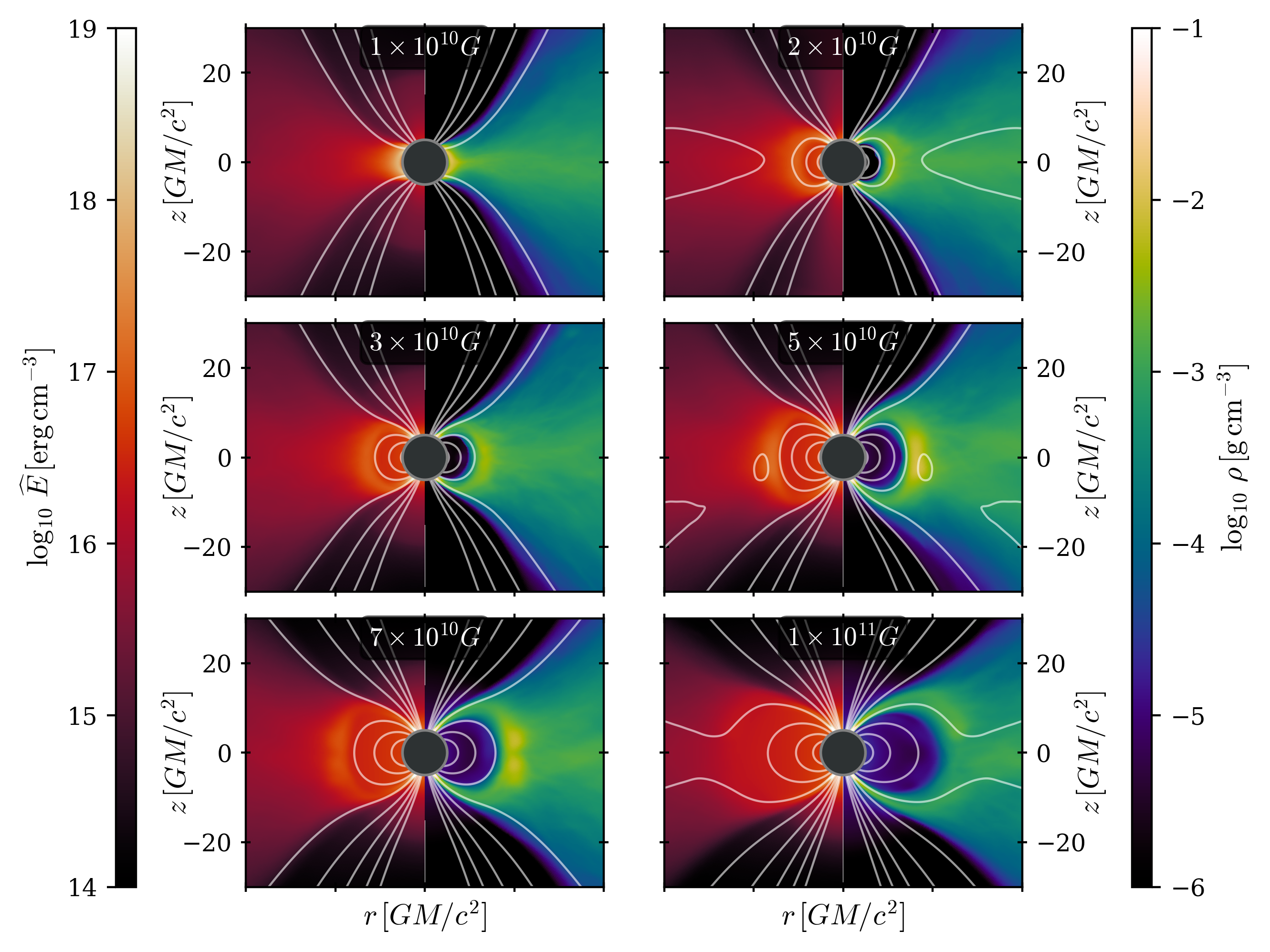}
    \caption{The radiation energy density $\hat{E}$ (\textit{left-half panel}) and the rest-mass density $\rho$ (\textit{right-half panel}) are shown in each panel. Magnetic field lines are plotted as iso contours of the vector potential, $A_{\phi}$. The magnetic dipole strength is as labeled on each frame. The plot is produced from the time-averaged data.}
    \label{magavg-grid}
\end{figure*}
\begin{figure}
    \centering
    \includegraphics[width = 0.9\linewidth] {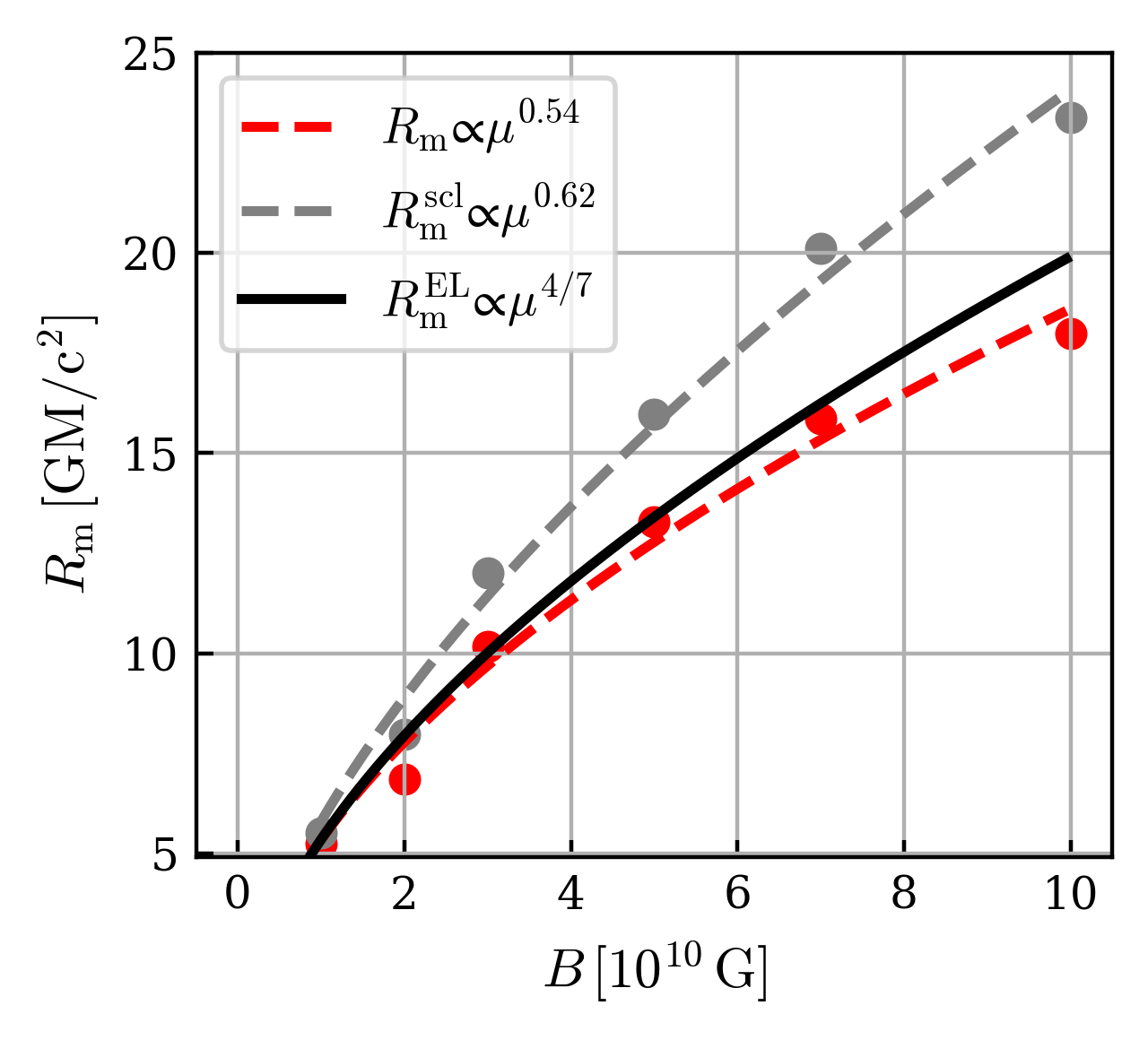}
    \caption{The magnetospheric radius $R_{\rm m}$ as a function of the dipole strength. The red circles are the computed radii in each magnetic field simulation. The grey circles mark the corresponding radii $R^{\rm scl}_{\rm m}$ of each magnetic field simulation scaled by $\mdot^{2/7}$ due to the different accretion rate in each simulation. The red and grey dashed lines show the fitted function on the computed data, and the black solid curve corresponds to the shape of the analytical solution of \cite{Elsner1977}.}
    \label{alfven_mag}
\end{figure}

The magnetic dipole strength and the accretion rate values affect the radius at which the disk is truncated. This influences how material follows the reconnected field lines onto the neutron star's surface, the accretion rate, and energy efficiency.
\paragraph{Magnetospheric radius} 
In accreting neutron stars with strong magnetic fields, the accretion disk is truncated where the ram pressure of the infalling matter is equal to the magnetic pressure.
At this radius, accreting material is forced to follow the magnetic field lines towards the neutron star's pole, forming accretion columns. 

The analytic magnetospheric radius, obtained by assuming the spherical accretion free-falling onto a dipolar magnetic field \citep{Elsner1977}, is

\begin{equation}\label{Alfven}
   R^{\rm EL}_{\rm m} = \left(\frac{\mu^4}{2GM\dot{M}^2}\right)^{1/7} = 3.2\times 10^8 \dot{M}^{-2/7}_{17}\mu^{4/7}_{30} \left(\frac{M}{M_\odot}\right)^{1/7} \, {\rm cm}.
\end{equation}
where the magnetic moment $\mu=\mu_{30} \cdot 10^{30}\,\mathrm{G\,cm}^3$, and the accretion rate  $\mdot=\mdot_{17} \cdot10^{17}\,$g/s.

From the simulations, we estimate the magnetospheric radius by the condition
\begin{equation}\label{alfcompute}
    (\rho + \frac{5}{3} u_{\rm int}) v^2 + \hat{E} = \frac{B^2}{2},
\end{equation}
where $\rho$ is rest-mass density, $u_{\rm int}$ internal energy, $\hat{E}$ radiation energy density and $B$ the magnetic dipole strength. 

\paragraph{Accretion rate} 
We compute the accretion $\mdot(r)$, inflow $\mdotin(r)$ and outflow $\mdotout(r)$ rates by integrating the momentum density $\rho u^{r}$ over the sphere at each radius of the accretion disk and accretion column, 
\begin{equation}\label{Eqmdot}
    \mdot (r) = -2\pi \int_0^\pi \rho u^r \sqrt{-g} d{\theta},
\end{equation}
\begin{equation}\label{Eqinflow}
    \mdotin (r) = -2\pi \int_{0}^\pi \rho u^r \sqrt{-g} d{\theta}\bigg|_{u^r<0},
\end{equation}
\begin{equation}\label{Eqoutflow}
    \mdotout(r)= 2\pi \int_{0}^\pi \rho u^r \sqrt{-g} d{\theta}\bigg|_{u^r>0}.
\end{equation}
The inflow rate is computed where $u^r$ is negative (flow is towards the accretor), and the outflow rate is in the opposite direction.
We can write $\mdot (r) = \mdotin (r) -  \mdotout (r)$, and $\sqrt{-g}= r^2 \sin \theta$, the square root of the metric determinant.
\paragraph{Energy flow} 
The energy flux is computed from the gas and radiation energy-momentum tensors, $T^r\phantom{}_t$ and $R^r\phantom{}_t$, respectively \citep[for more details on energy flow equations, see][]{sadowskienergyflow}. The total flux $T^r\phantom{}_t+R^r\phantom{}_t$ is not an interesting quantity from the observational point of view, because the rest-mass density is included in $T^r\phantom{}_t$. We add $\rho u^r$ to the total energy flux to remove the rest-mass contribution.\footnote{With the metric signature (-,+,+,+); lowering time index of tensors introduces a negative sign.} The total and radiation fluxes are
\begin{equation}\label{totalflux}
    F^r_{\rm tot} = -(T^r\phantom{}_t + \rho u^r) - R^r\phantom{}_t,\quad F^r_{\rm rad} = -R^r\phantom{}_t,
\end{equation}
respectively. $F^r_{\rm rad}$ reflects the energy carried by radiation which either propagates freely or advects with the gas. The plasma energy flux $T^r\phantom{}_t+\rho u^r$ can be decomposed into different forms of energy flux as
\begin{equation}\label{energymomentumtensor}
     T^r\phantom{}_t + \rho u^r = 
      \gamma u_{\rm int} u^r u_t + (b^2 u^r u_t - b^r b_t) + \rho u^r (1+u_t),
\end{equation}
where the internal energy is $u_{\rm int} = p/(\gamma - 1)$. 
Here $p$ is the gas pressure, $\gamma = 5/3$ is the adiabatic index, $u^\mu$ is the velocity 4-vector and $b^\mu$ is the magnetic 4-vector as detailed in Eq.~\ref{4beq}. Eq.~\ref{energymomentumtensor} is split into the internal energy and the magnetic energy fluxes,
%
\begin{equation}\label{internal}
    F^r_{\rm int} = - \gamma u_{\rm int} u^r u_t,\quad F^r_{\rm mag} = - (b^2 u^r u_t -  b^r b_t)
\end{equation}
%
%
%
respectively, and the remaining term in Eq.~\ref{totalflux} is the binding energy flux \citep{sadowski2016}: 
\begin{equation}\label{bindingflux}
    F^r_{\rm bind} = - \rho u^r (1+u_t).
\end{equation}
It contains information about the gravitational and kinetic energy fluxes
\begin{equation}\label{gravflux}
    F^r_{\rm grav} = - \rho u^r (1-\sqrt{-g_{tt}}),\quad F^r_{\rm ke} = - \rho u^r (u_t+\sqrt{-g_{tt}}).
\end{equation}
The luminosity of each flux component is obtained by integrating the corresponding flux over the sphere at each radius: 
\begin{equation}\label{energylum}
    L_{\rm EN} =  2\pi \int_{0}^\pi F^r_{\rm EN} \sqrt{-g} d{\theta},
\end{equation}
where the form of energy flux is defined by index ${\rm EN}$.

In the following subsections, we present the parameters calculated from the time-averaged data for each simulation, organised according to the studies by magnetic dipole strengths and accretion rates.

\subsection{Study of magnetic dipole strengths}

\begin{figure}
    \centering
    \includegraphics[width =\linewidth]{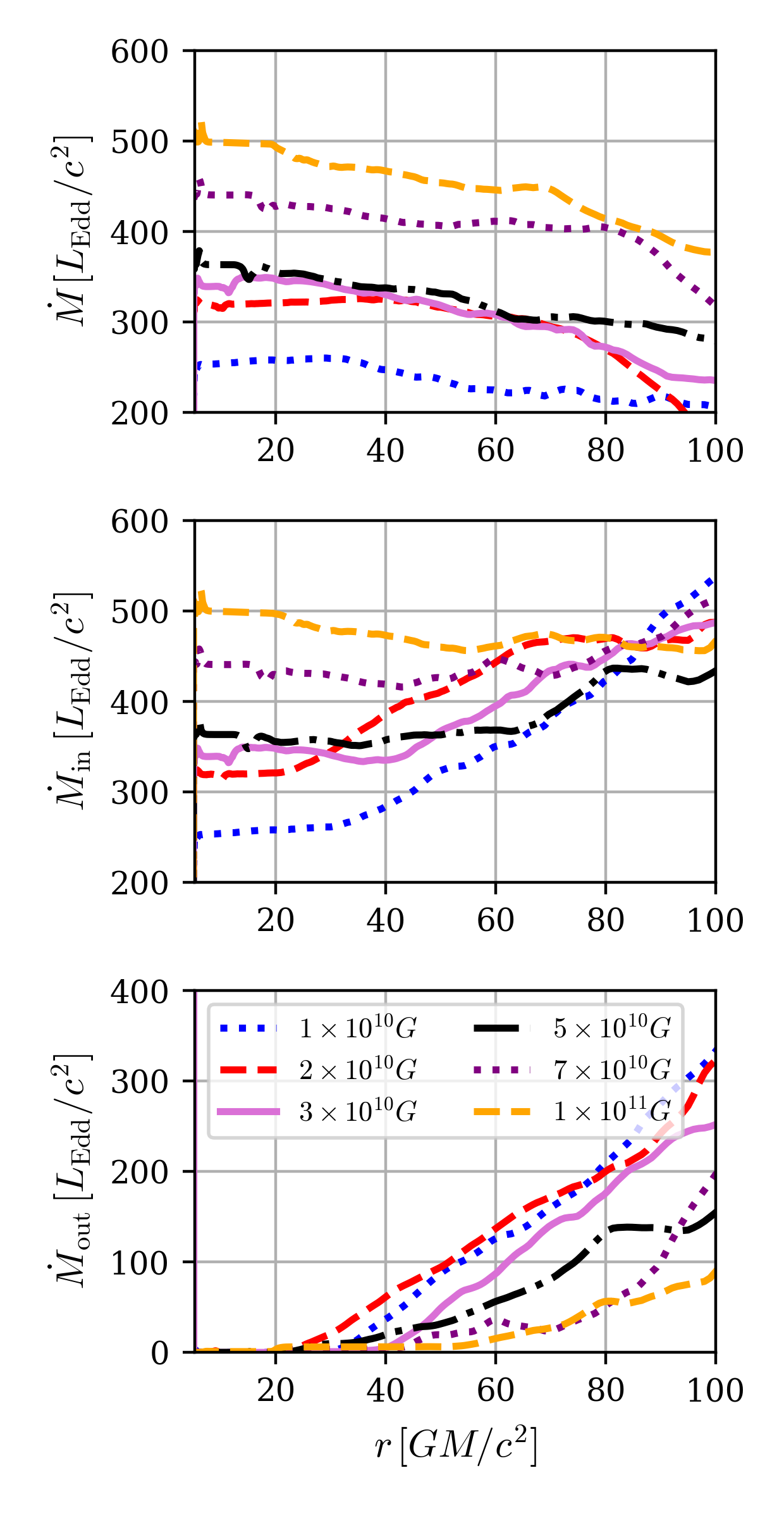}
    \caption{The accretion $\mdot$, inflow $\mdotin$, and outflow $\mdotout$ rates in the unit of $L_{\rm Edd}/c^2$ are shown in the top, middle, and bottom panels, respectively. Simulations with different strength of magnetic dipoles are shown in different colors and line styles.}
    \label{mdot_mag}
\end{figure}
\begin{figure}
    \centering
    \includegraphics[width =\linewidth]{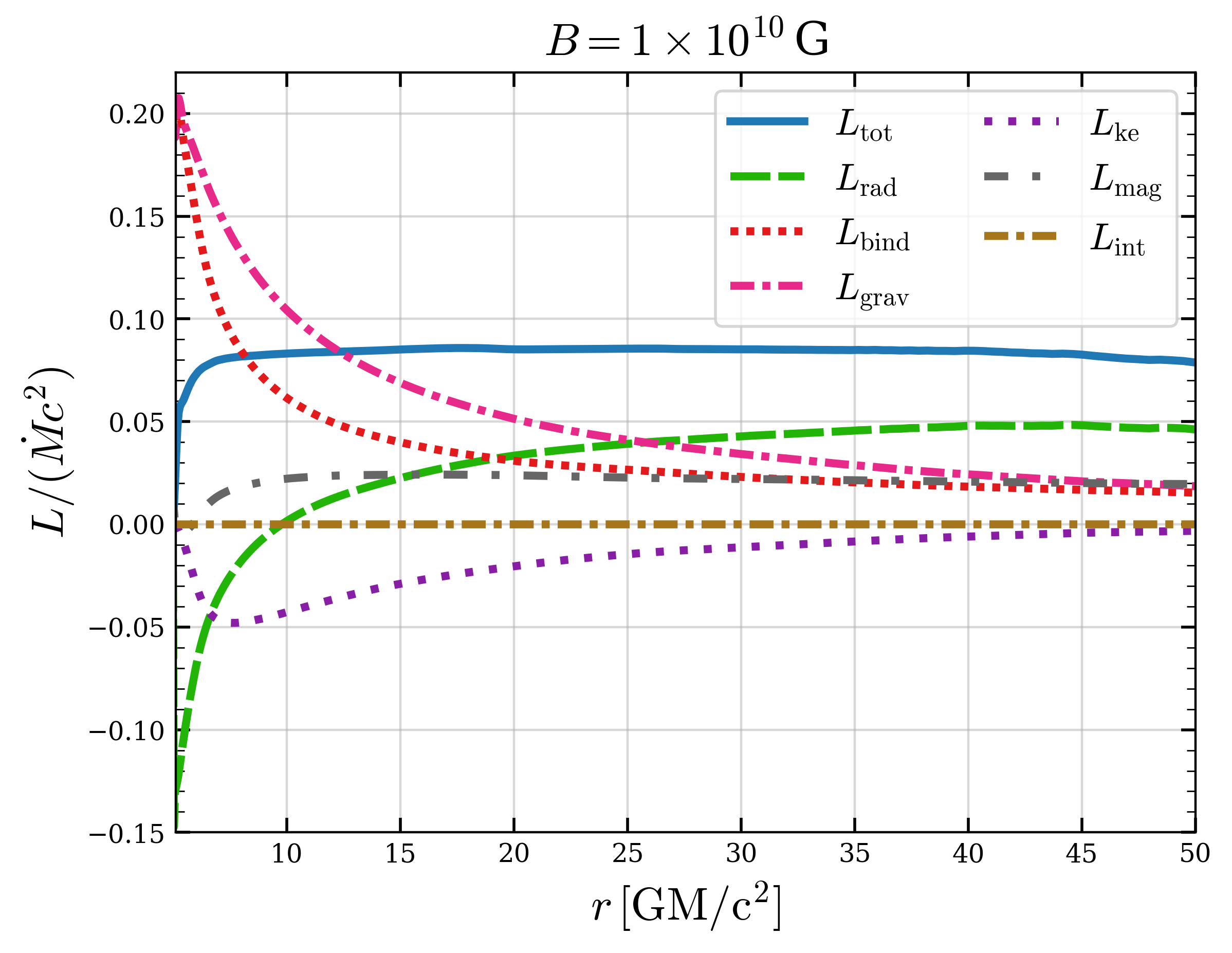}
    \includegraphics[width =\linewidth]{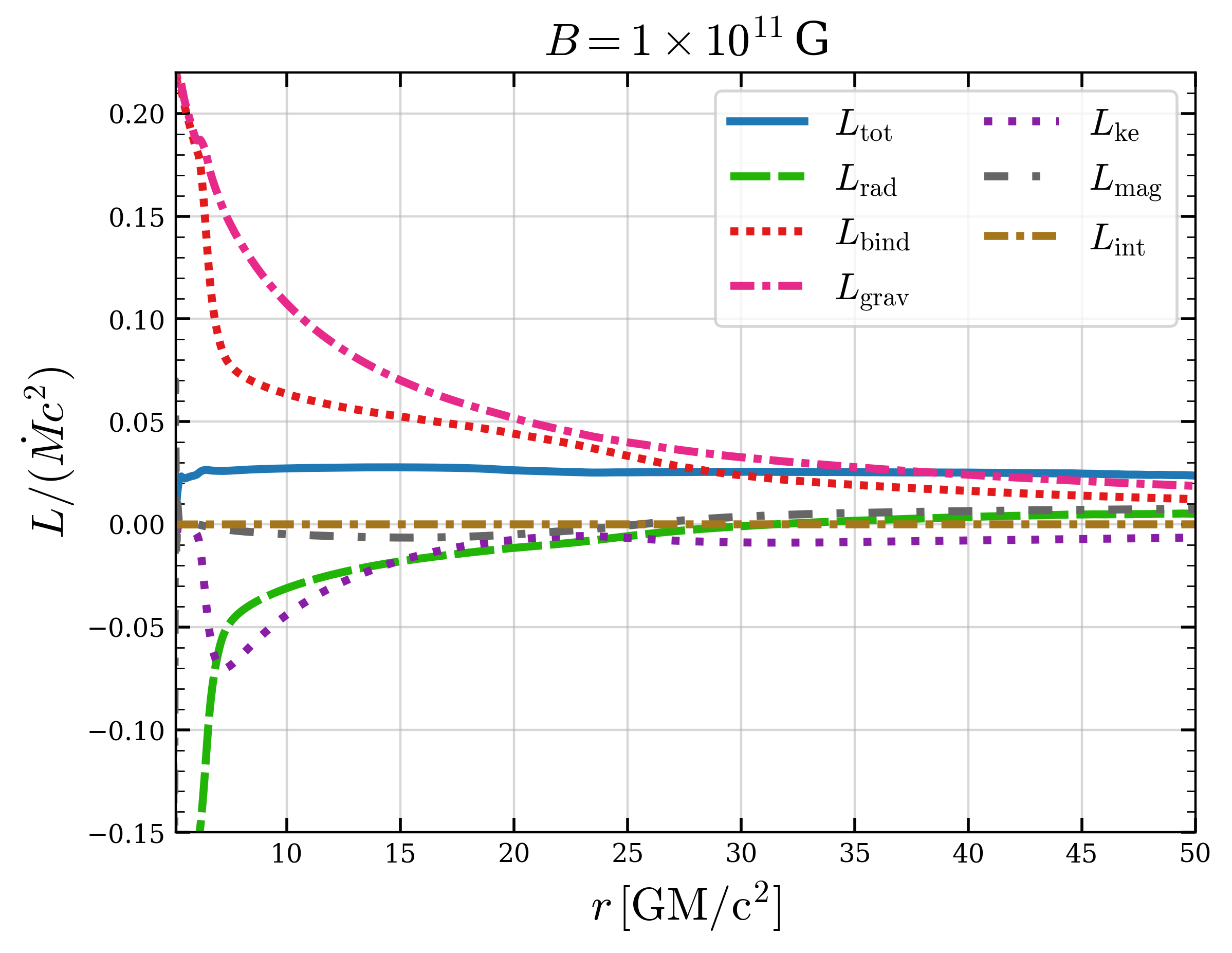}
    \caption{The efficiency $L/(\dot{M}c^2)$ of total luminosity $L_{\rm tot}$ and its components for two simulations with dipole $10^{10}\,$G (\textit{top panel}) and $10^{11}\,$G (\textit{bottom panel}). Efficiency in each energy is denoted by a different color and line style.}
    \label{energybug}
\end{figure}
\begin{figure*}
    \centering
    \includegraphics[width = 0.75\linewidth]{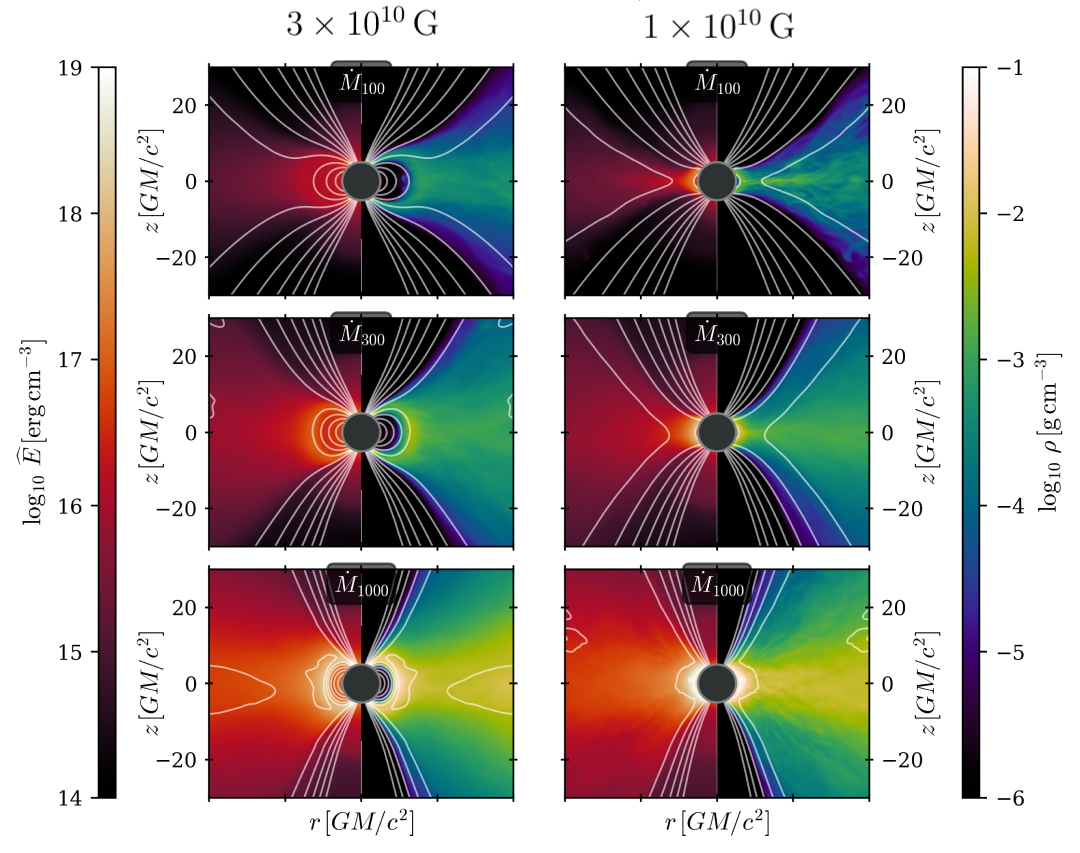}
    \caption{The radiation energy density $\hat{E}$ (\textit{left-half panel}) and the rest-mass density $\rho$ (\textit{right-half panel}) are shown in each panel. Magnetic field lines are plotted as contours of the vector potential, $A_{\phi}$. The left column shows the simulation with $3\times10^{10}\,$G and the right column with $1\times10^{10}\,$G, with the accretion rate labeled on each frame.}
    \label{avgmdotgtid}
\end{figure*}
\begin{figure}
    \centering
    \includegraphics[width = 0.8\linewidth]{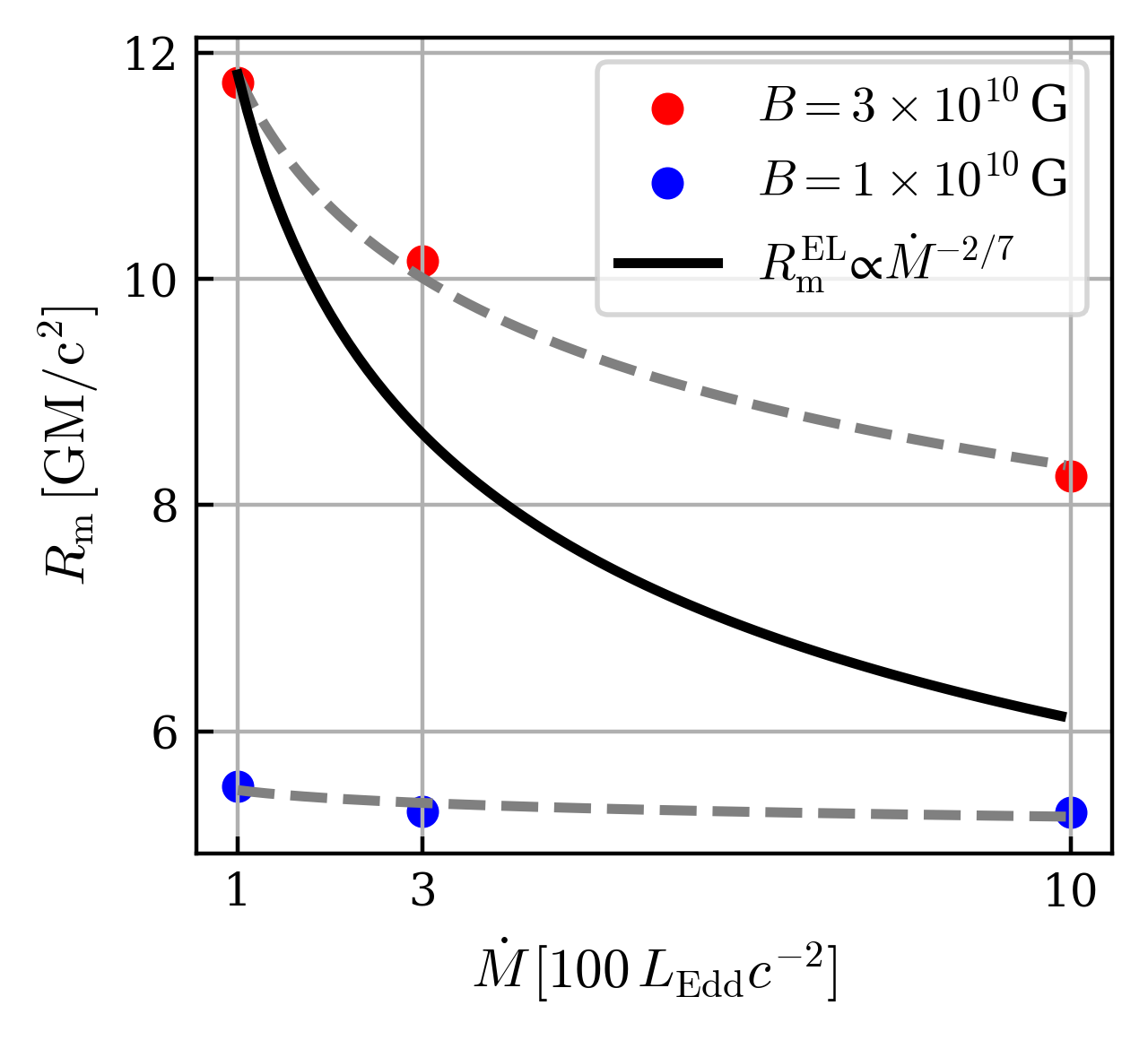}
    \caption{Magnetospheric radius $R_{\rm m}$ versus accretion rate in simulations with two different magnetic dipole strengths. Each color corresponds to the simulations with a given magnetic dipole strength. The dashed grey lines represent fitted functions, while the solid black line denotes the analytical solution of \cite{Elsner1977}, multiplied by a suitable constant}.
    \label{alfven_mdot}
\end{figure}
%
\begin{figure}
    \centering
    \includegraphics[width =\linewidth]{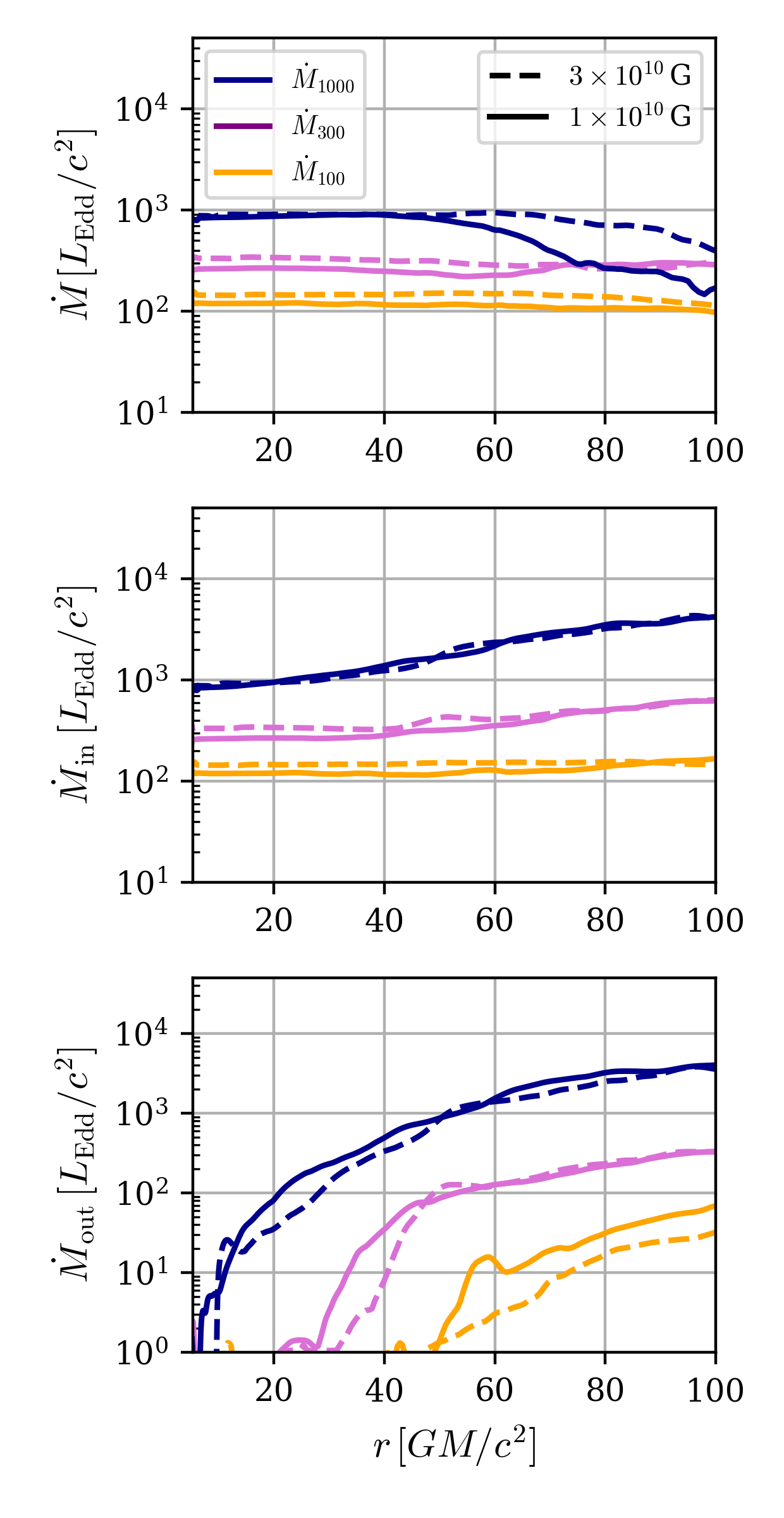}
    \caption{The accretion $\mdot$, inflow $\mdotin$, and outflow $\mdotout$ rates in the unit of $L_{\rm Edd}/c^2$ are shown in the top, middle, and bottom panels, respectively. The line styles represent different magnetic dipole strengths, and the colors represent different accretion rates.}
    \label{mdotmdot}
\end{figure}
\begin{figure}
    \centering
    \includegraphics[width =\linewidth]{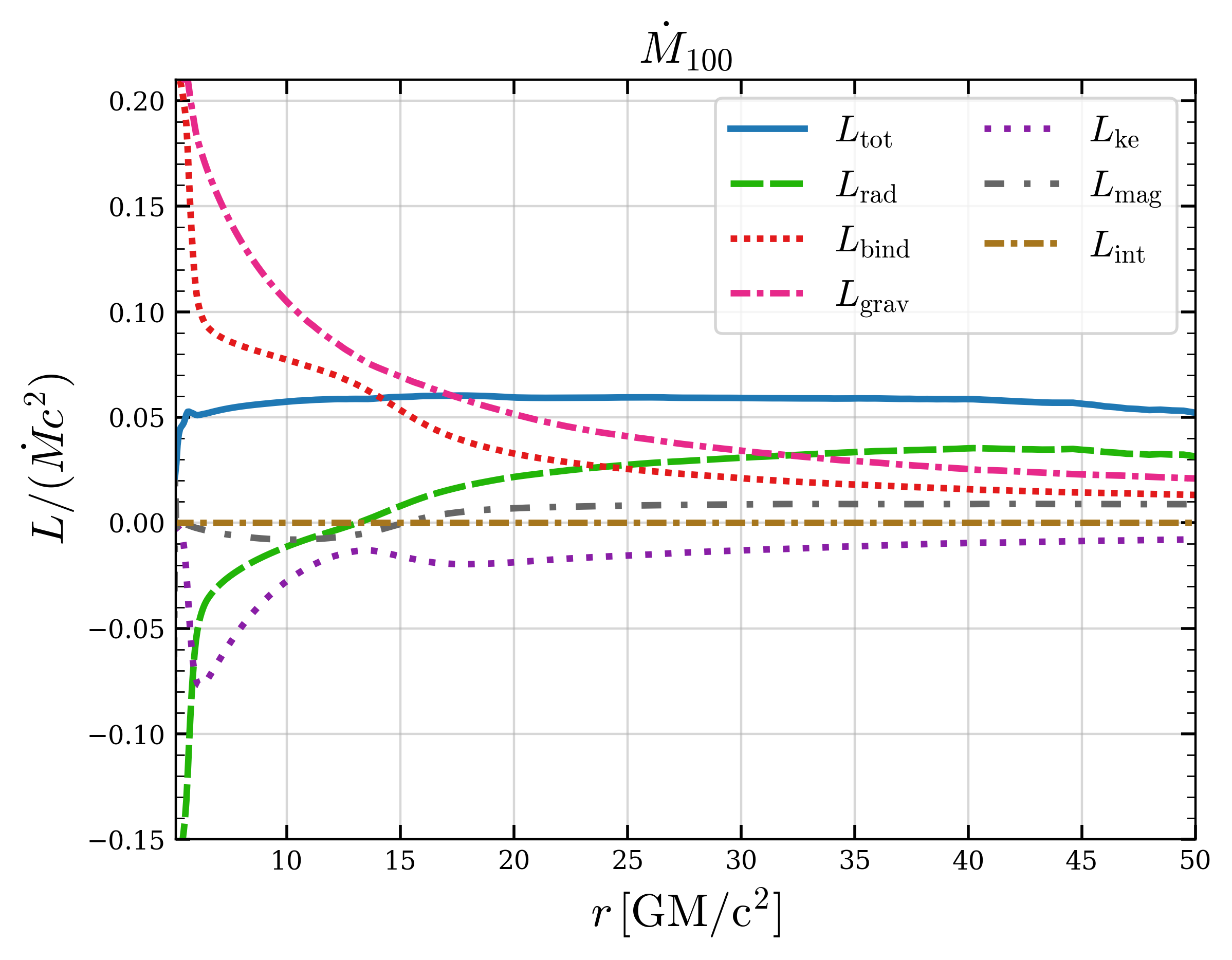}
    \includegraphics[width =\linewidth]{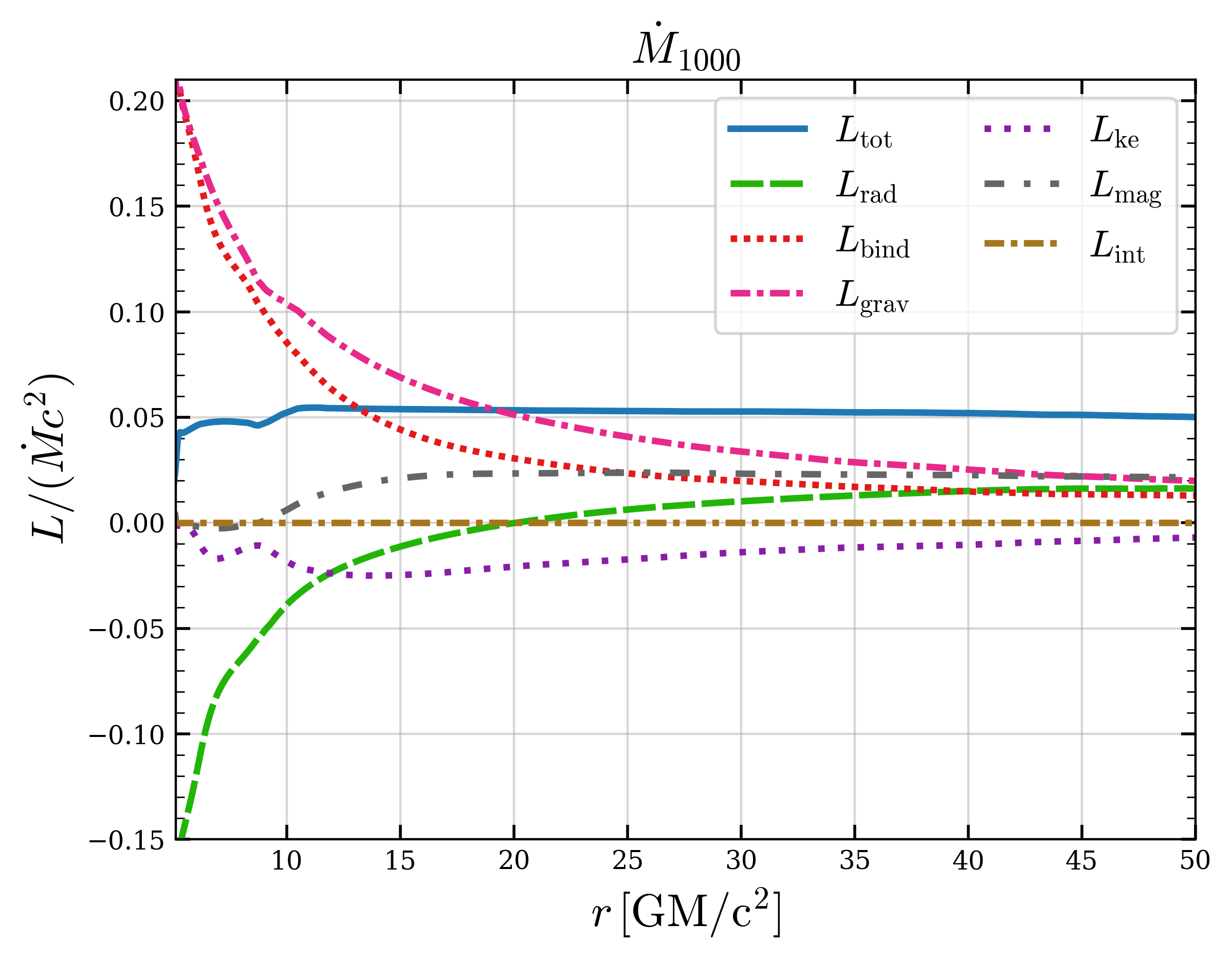}
    \caption{The efficiency $L/(\dot{M}c^2)$ of total luminosity $L_{\rm tot}$ and its components for two simulations with accretion rate of $\mdot_{100}$ (\textit{top}) and $\mdot_{1000}$ (\textit{bottom}). Each efficiency is denoted by a different color and line style.}
    \label{energybug2}
\end{figure}
In Fig.~\ref{magavg-grid}, we show the time-averaged data of six simulations with different magnetic dipole strengths that are labeled in each frame. The first observed trend is the change in the accretion disk truncation radius relative to the magnetic dipole strength. In the simulation with a weak dipole $10^{10}\,$G in the top left panel, nearly all dipole lines break, allowing matter to approach the neutron star's surface in the equatorial region. As the dipole strength increases, magnetic reconnection occurs further from the star. The dipole lines near the star remain unchanged, and matter channelled along these lines falls onto the neutron star's surface near the polar region. 

From Fig.~\ref{magavg-grid} it follows that accreting material propagates to the higher latitudes for reduced strength of the magnetic dipole. Consequently, the radiation energy $\hat{E}$ in the left-half panels is accumulated in a narrower region near the polar axis in the weaker magnetic dipole simulation. 
 
The magnetospheric radii $R_{\rm m}$, estimated using Eq.~\ref{alfcompute} and the time-averaged data for the simulations with different magnetic dipole strengths, are shown with red circles in Fig.~\ref{alfven_mag}.
Following this section, we show that the accretion rate gradually increases with the strength of the dipole of the neutron star. We compute the scaled value of the magnetospheric radius with the accretion rate based on the analytical solution (see Eq.~\ref{Alfven} where the magnetospheric radius is proportional to $\mdot^{-2/7}$). The grey circles in Fig.~\ref{alfven_mag} are the scaled values $R^{\rm scl}_{\rm m} = R_{\rm m} \mdot^{2/7}$ where $\mdot$ is accretion rate in the units of Eddington luminosities. The red and grey dashed lines are the fitted functions on computed values $R_{\rm m}$ and $R^{\rm scl}_{\rm m}$, respectively. The black solid line shows, $R^{\rm EL}_{\rm m}$, corresponding to the analytical solution of \cite{Elsner1977} (see Eq.~\ref{Alfven}), with an arbitrary normalization. 
We find that $R_{\rm m}$ as a function of the magnetic field follows the analytical solution, obtained by assuming the spherical accretion free-falling onto a dipolar magnetic field \citep{Elsner1977} with $R^{\rm EL}_{\rm m}\propto \dot{M}^{-2/7}$ 
to a 20\%. 
In our simulation, the accretion rate is high, and the additional source of radiation pressure in the inner part of the accretion disk influences the magnetospheric radius.

The accretion, inflow, and outflow rates computed using Eqs.~\ref{Eqmdot}-\ref{Eqoutflow}, are depicted in the top, middle, and bottom panels of 
Fig.~\ref{mdot_mag}, respectively.

The results generated from time-averaged data indicate that for radii less than $50\, r_{\rm g}$ the accretion rate $\mdot(r)$ remains approximately constant across nearly all magnetic field strengths. As the magnetic field strength increases, the outer radius of the region of uniform accretion rate shifts closer to the star, likely due to the longer convergence time required for stronger dipole field simulations, as the disk takes more time to reach stability. 
However, to enable comparison, in this study, we use the same time-averaging duration for all presented simulations. 

The results show that the accretion rate increases with the dipole strength. In the presence of a strong dipole, the accreting material is channeled along the field lines, leading to the accumulation of most of the material in the inner radii, which suppresses the outflow. As the magnetic field weakens, outflows become more prominent, with a thick outflow forming at radii beyond $40\, r_{\rm g}$ and gradually extending outward. Near the neutron star surface, the outflow diminishes to zero in all simulations.

The efficiency of each energy component is computed using Eq.~\ref{energylum} and the time-averaged data. To compare the energy flow across different models, we compute the energy efficiency as the ratio $L_{\rm EN}/(\mdot c^2)$. The accretion rate $\mdot$ is computed at the radius $20\, r_{\rm g}$, where the accretion rate in all simulations remains constant. 

In Figure~\ref{energybug} is presented the dimensionless energy efficiency for each energy component, along with the total energy efficiency. For a dipole strength of $10^{10}\,$G, the total energy efficiency reaches approximately $0.08$, whereas for the stronger dipole field of $10^{11}\,$G, it declines to about $0.03$. In our simulations, stronger magnetic dipoles reduce the overall accretion efficiency.

In the weak dipole case, most magnetic field lines are open, allowing material to accrete onto the stellar surface through the equatorial region. With the open field, the turbulent accretion can expand toward higher latitudes. In contrast, for the strong dipole configuration, magnetic reconnection occurs at the radius $R_{\rm m} \approx 18\,r_{\rm g}$, channeling accretion flow along the remaining closed field lines toward the polar regions. In such a configuration, accretion is restricted to low latitudes, resulting in a less turbulent flow structure. Consequently, the contributions of radiation $L_{\rm rad}$ and kinetic $L_{\rm kin}$ luminosities are lower in the strong dipole than in the weak dipole case.

As material accretes onto the central object, it releases binding energy, while gas located closer to the star becomes more gravitationally bound. The binding luminosity, $L_{\rm bind}$, is shown in Fig.~\ref{energybug} together with the gravitational luminosity $L_{\rm grav}$. 

The radiation luminosity $L_{\rm rad}$ becomes negative near the neutron star, 
since most of the radiation is carried inward by the flow—a consequence of the well-known photon-trapping effect \citep[see][]{Ohsuga2002, SadowskiNarayan2016}. However, for an accreting neutron star, this advected radiation is expected to be ultimately emitted from the stellar surface. 
Following \citet{abarca+21}, we implement energy-reflective boundary conditions at the neutron star surface. 
Still, this method does not fully reproduce the expected behavior. A more detailed analysis of the radiation luminosity is presented in Section~\ref{Sapparent}.

In Fig.~\ref{energybug}, the kinetic efficiency is negative at radii close to the star, indicating that the kinetic energy associated with Keplerian motion is advected inward by the infalling gas. As the radius increases, the kinetic efficiency approaches zero, where the inward advection and outward release of kinetic energy are nearly balanced. 
If we extend the curves to the larger radii, the kinetic efficiency becomes positive. This would indicate substantial outflows occurring at $r\geq 50\, r_{\rm g}$ in the simulation with the weak dipole and at even greater distances in the simulations with the strong dipole (see Fig.~\ref{mdot_mag}).
 
The efficiency associated with the magnetic luminosity $L_{\rm mag}$ is shown in Fig.~\ref{energybug}. Magnetic energy facilitates the outward transport of released energy from the innermost regions of the disk and accretion column, redistributing it toward the outer regions. 
In the weak magnetic dipole simulation, the magnetic luminosity exceeds that observed in simulations with stronger magnetic fields. 
 With a strong magnetic dipole, the ingoing and outgoing magnetic fluxes are balanced, leading to a net magnetic luminosity that approaches zero. Since radiation cools the accretion disk in our simulations, the internal energy, $L_{\rm int}$ 
 is approximately zero in both simulations.
 
\subsection{Study of accretion rates}
In Fig.~\ref{avgmdotgtid} are shown the time-averaged data of three simulations with different accretion rates for two different magnetic dipole strengths. For each magnetic dipole strength, increasing the accretion rate decreases the radius at which the disk is truncated (top to bottom panels in Fig.~\ref{avgmdotgtid}). Consequently, the accretion expanded to higher latitudes with increasing the accretion rate. 

In Fig.~\ref{alfven_mdot} we show the magnetospheric radius versus the accretion rate in the unit of $100\,L_{\rm Edd}/c^2$. We estimate the magnetospheric radius $R_{\rm m}$ using Eq.~\ref{alfcompute}.

With each magnetic dipole strength
simulation, the magnetospheric radius, $R_{\rm m}$, decreases with increasing accretion rate. 
In the cases with weak magnetic field ($10^{10}\,$G), the magnetosphere is located very close to the neutron star surface and cannot recede further as the accretion rate increases, so that $R_{\rm m}$ remains nearly constant over our range of accretion rates.
Even for the stronger magnetic fields, the decline in $R_{\rm m}$ is not as steep as the analytical Elsner-Lamb prediction $R^{\rm EL}_{\rm m}$. This might be caused by the additional pressure exerted by radiation, which becomes increasingly significant at higher accretion rates. 

The radial profiles of mass fluxes in the different flow components are presented in Fig.~\ref{mdotmdot}.
The outflow rate significantly increases with the accretion rate. As the accretion rate decreases, the radial location of the outflows shifts outwards because of the larger magnetospheric radius at lower accretion rates.

 In Figure~\ref{energybug2} we show the energy efficiency in simulations with varying accretion rates for the simulations with two magnetic dipole strengths. 

Within the explored accretion rate range, the total efficiency of the accreting system is largely independent of the accretion rate. The distribution of energy efficiency among the components varies with accretion rate. The radiation efficiency $L_{\rm rad}/\mdot c^2$ is lower in the high accretion rate simulation. Conversely, the kinetic efficiency $L_{\rm ke}/\mdot c^2$ near the neutron star is more negative in case with the low accretion rate compared to the high accretion rate simulations. The outflows are more powerful in the high accretion rate scenario. The magnetic efficiency $L_{\rm mag}/\mdot c^2$, which is responsible for transporting and redistributing energy at larger radii, is larger in simulations with larger accretion rates.

\section{Power of outflow and apparent luminosity}\label{Sapparent}

\begin{figure*}
    \centering
    \includegraphics[width =  \linewidth]{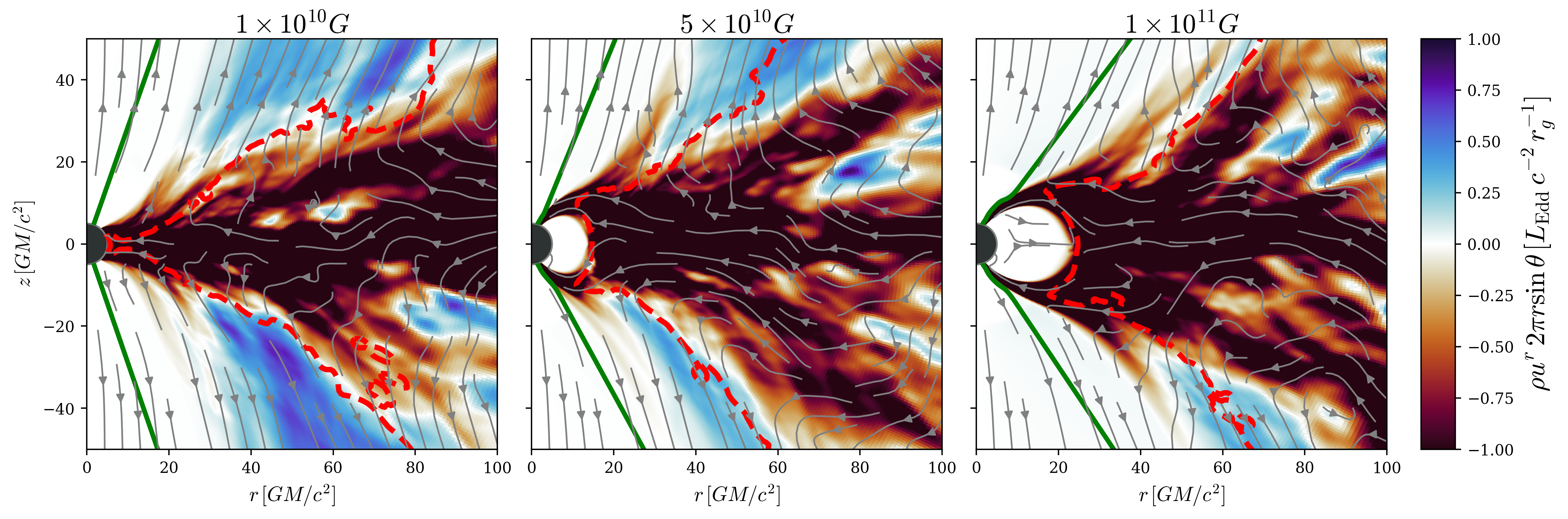}
    \caption{The momentum density $\rho u^r$, multiplied by $2\pi r \sin \theta$, is expressed in units of $[L_{\rm Edd}c^{-2} r_{\rm g}^{-1}]$. Negative values indicate inflow, while positive values correspond to outflow. The red dashed line represents the zero Bernoulli surface $Be = 0$, and the green solid line is the photosphere $\tau_{\rm r} = 1$. The arrows depict the radiation flux direction, and the magnetic field strength is labeled above each frame.}
    \label{outflow}
\end{figure*}
\begin{figure}
    \centering
    \includegraphics[width = \linewidth]{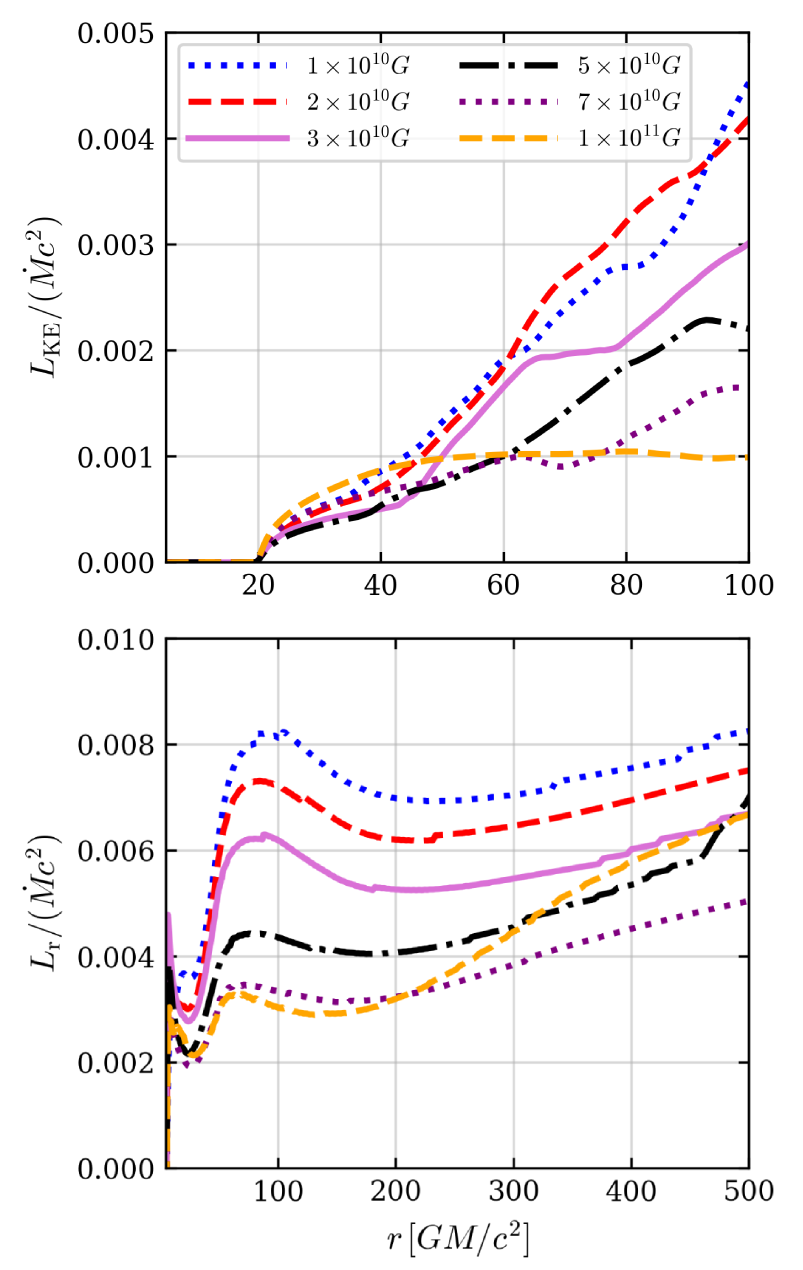}
    \caption{The efficiency of outgoing kinetic luminosity $L_{\rm KE}/(\mdot c^2)$ computed in outflow region with $Be > 0$, $u^r>0$ (\textit{top panel}) and the efficiency of radiation luminosity $L_{\rm r}/(\mdot c^2)$ computed in the optically thin region with $\tau_r<1$ (\textit{bottom panel}). The strength of the dipole in each simulation is shown in a different color and line style.}
    \label{ke_rad_efficiency_mag}
\end{figure}

\begin{figure}
    \centering
    \includegraphics[width =0.70\linewidth]{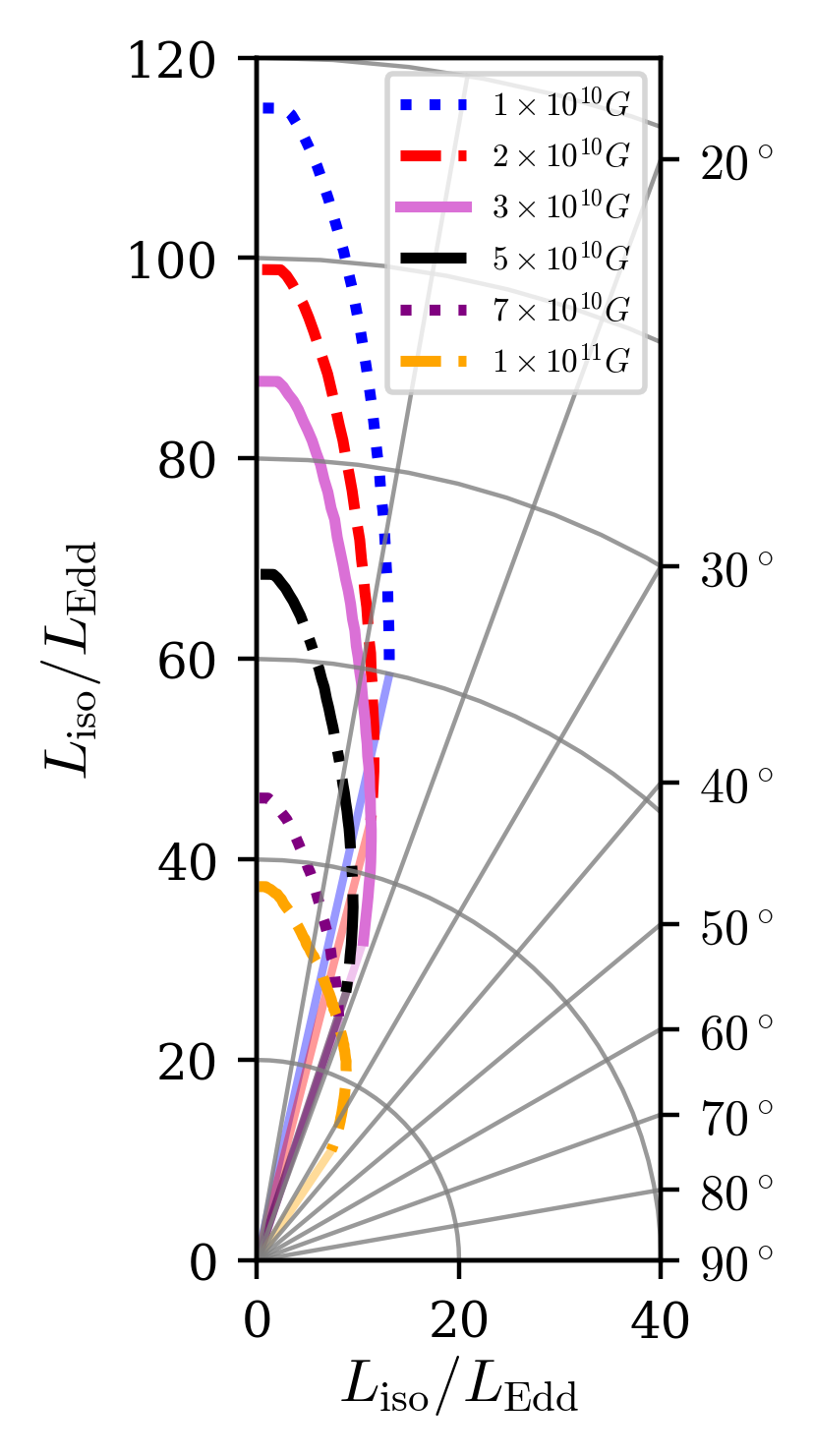}
    \caption{Isotropic luminosity in the unit of $L_{\rm Edd}$ in polar coordinates. The straight grey solid lines correspond to particular viewing angles. The right side labels show viewing angles. The faded colored straight line in each magnetic field curve represents the optically thick region at which the radiation flux diminishes to zero.}
    \label{apparent_mag}
\end{figure}

%
\begin{figure*}
    \centering
    \includegraphics[width = \linewidth]{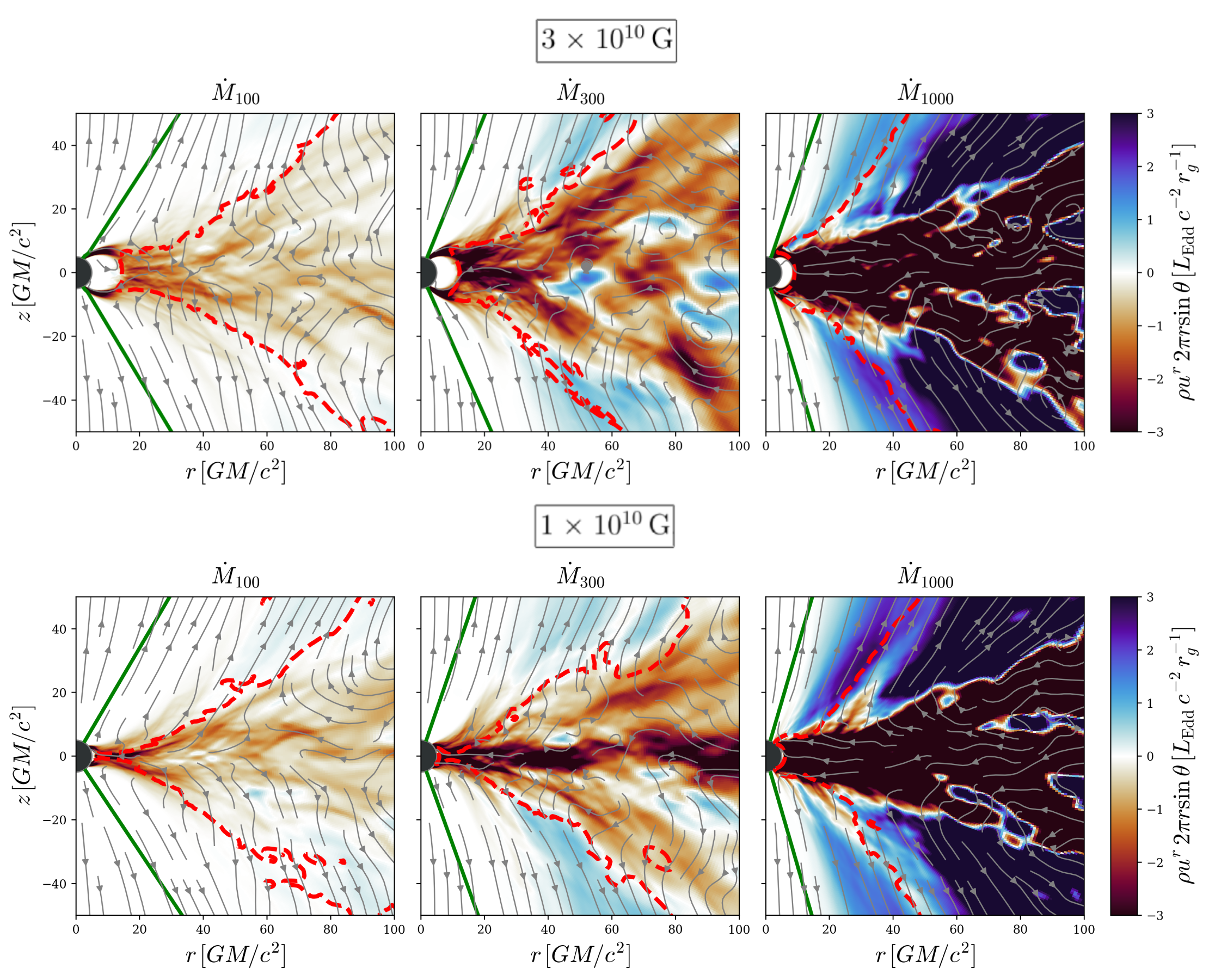}
    \caption{The momentum density $\rho u^r$, multiplied by $2\pi r \sin \theta$, is expressed in units of $[L_{\rm Edd}c^{-2} r_{\rm g}^{-1}]$. Negative values indicate inflow, while positive values correspond to outflow. The red dashed line represents the zero Bernoulli surface $Be = 0$, and the green solid line is the photosphere $\tau_{\rm r} = 1$. The top and bottom rows correspond to the magnetic dipole strengths of  $3\times 10^{10}\,$G and $10^{10}\,$G. Streamlines depict the radiation flux direction, and the accretion rate is labeled above each frame.}
    \label{outflowmdot}
\end{figure*}
\begin{figure}
    \centering
    \includegraphics[width = \linewidth]{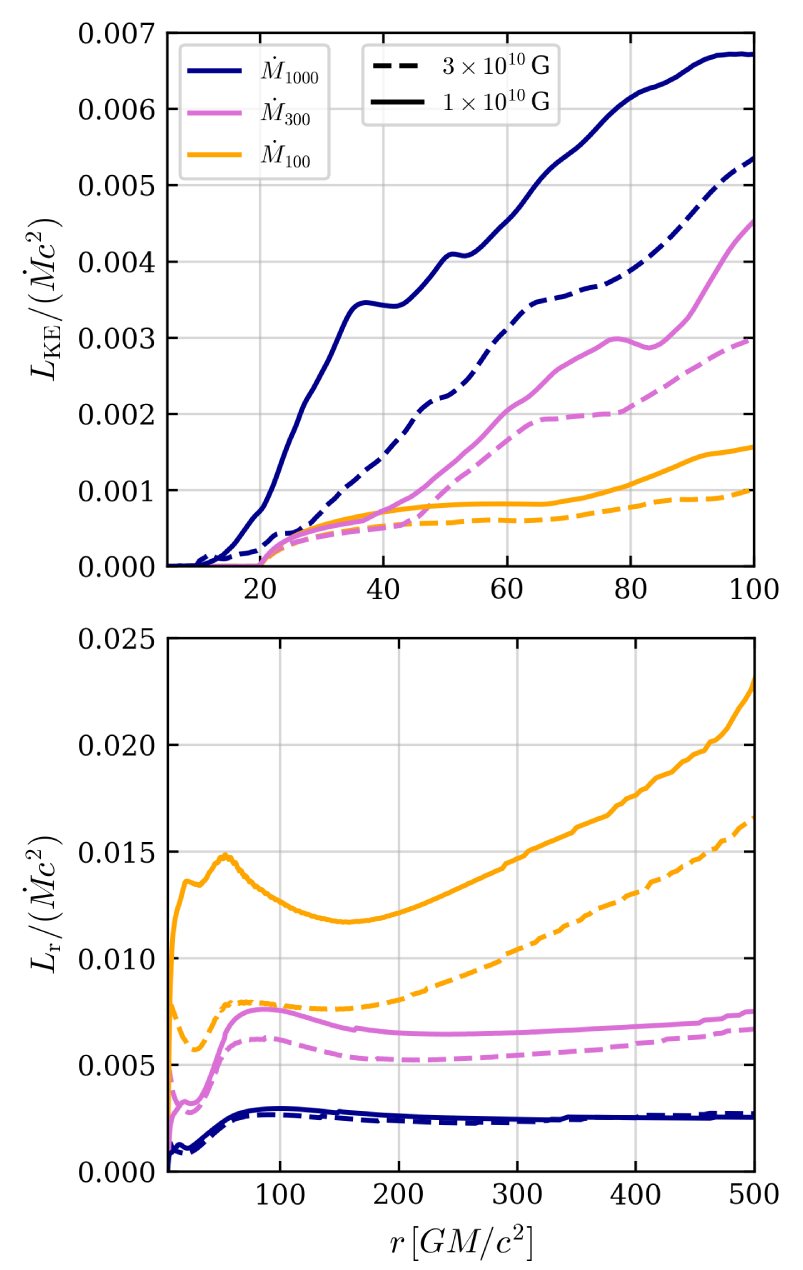}
    \caption{The efficiency of outgoing kinetic luminosity $L_{\rm KE}/(\mdot c^2)$ computed in outflow region with $Be > 0$, $u^r>0$ (\textit{top panel}) and the efficiency of radiation luminosity $L_{\rm r}/(\mdot c^2)$ computed in the optically thin region with $\tau_r<1$ (\textit{bottom panel}). The accretion in each simulation is shown in a different color, and the magnetic dipole strength is shown in line style.}
    \label{ke_rad_efficiency_mdot}
\end{figure}
%
\begin{figure}
    \centering
    \includegraphics[width =0.7 \linewidth]{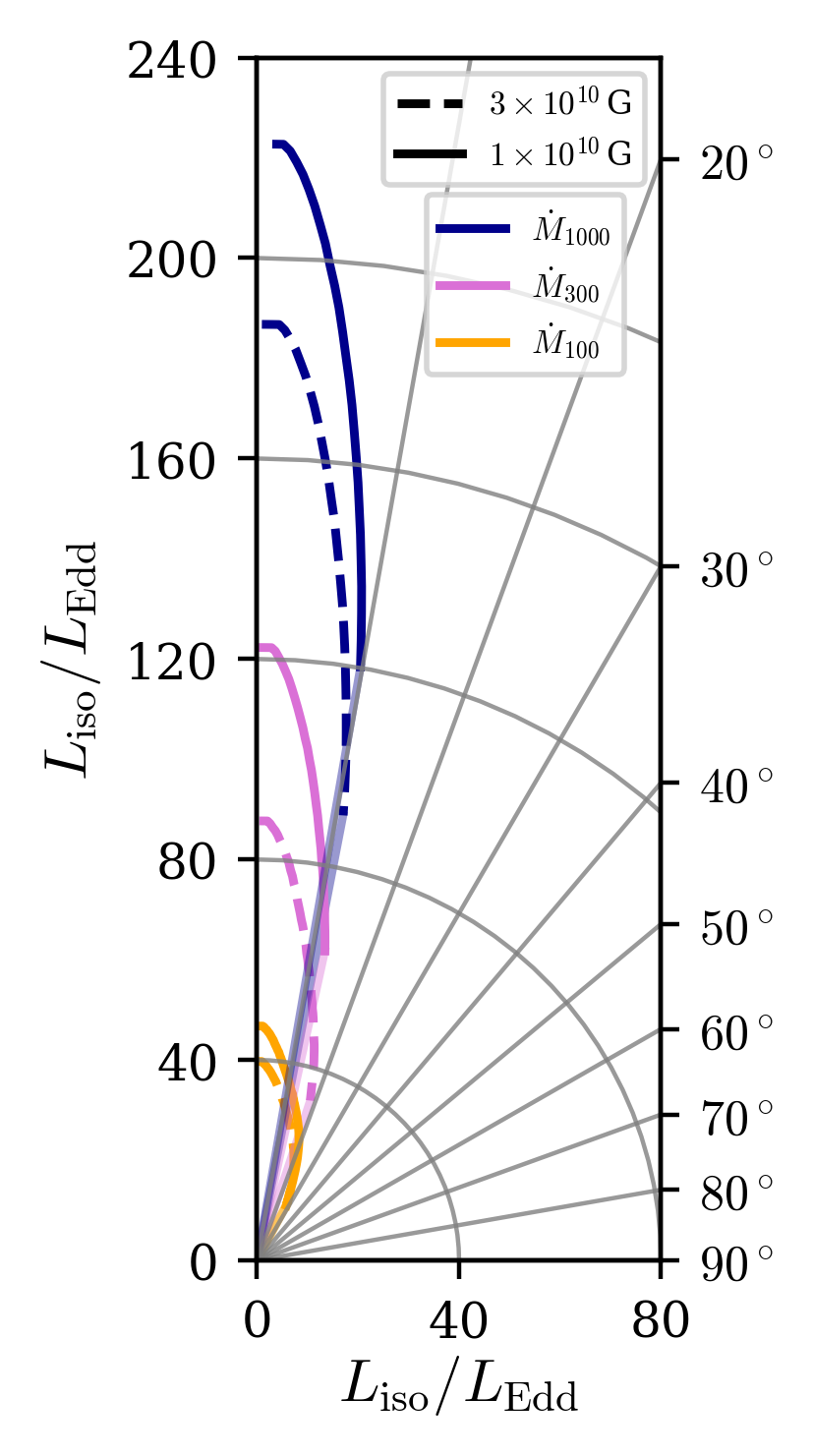}
    \caption{The isotropic luminosity, expressed in units of $L_{\rm Edd}$, is displayed in polar coordinates. The right side labels indicate the corresponding viewing angles. Straight grey solid lines denote specific viewing angles. For each magnetic field and accretion rate, the faded portion of the curve represents the optically thick region, where the radiation flux drops to zero.}
    \label{apparmdot}
\end{figure}
\begin{figure*}
    \centering
    \includegraphics[width=0.5\linewidth]{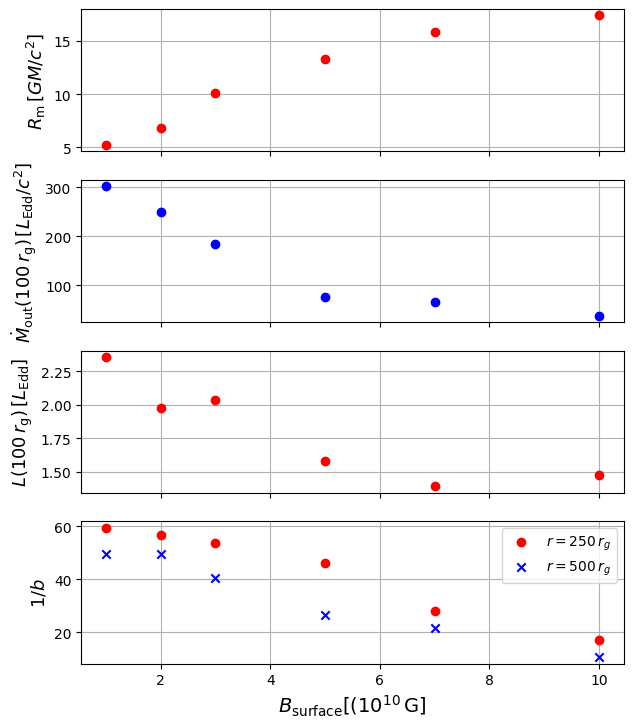}\hfill
    \includegraphics[width=0.5\linewidth]{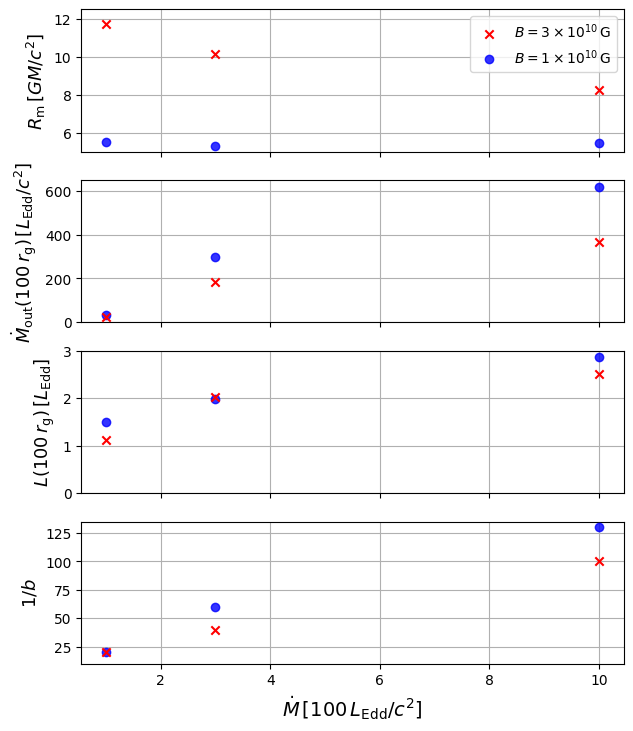}
    \caption{The parameters extracted from the time-averaged data of the magnetic dipole study (\textit{left panel}) and the accretion rate study (\textit{right panel}). The parameters are magnetospheric radius $R_{\rm m}$, outflow rate $\mdotout$, luminosity $L$ (computed in the optically thin region) and the beaming factor, which is shown as $1/b$, from top to bottom panels, respectively. The beaming factor is computed at two radii 250~$r_{\rm g}$ and 500~$r_{\rm g}$ in the magnetic dipole study and at 500~$r_{\rm g}$ in the accretion rate study and the outflow rate and luminosity are computed at radius 100~$r_{\rm g}$. In the accretion rate study (\textit{right}), the parameters are shown for two magnetic dipoles, the red x markers and the blue circles denote the magnetic dipole $3\times 10^{10}\,$G and $1\times 10^{10}\,$G, respectively.}
    \label{conclusions}
\end{figure*}

In this section we revisit the key parameters of ULXs, including the efficiencies of kinetic and radiation luminosities as well as the apparent luminosity. 

Super-Eddington accretion rates produce intense radiation in the region between the magnetospheric radius and the spherization radius, which in turn drives a strong outflow within this zone \citep{Sunyaev1973}. This outflow effectively collimates radiation along the polar axis. In our simulations, the spherization radius is estimated using $R_{\rm sph} \approx 15\,\dot{m}\,r_{\rm g}$ where $\dot{m} = \dot{M}/\dot{M}_{\rm Edd}$ with the accretion rate computed in the region of uniform accretion, and lies in the range of approximately $240$–$500\,r_{\rm g}$. 

\subsection{Study of magnetic dipole strengths} 
In Fig~\ref{outflow}, we show the inflow-outflow map of simulations with three different magnetic dipole strengths. The colormap shows the momentum density $\rho u^r$ multiplied by $2\pi r \sin\theta$. The negative and positive values indicate the inflow and outflow, respectively. The grey streamlines show the direction of the radiation flux, the green solid line represents the photosphere, defined as the surface where the optical depth satisfies $\tau_{\rm r} = 1$, as discussed later in this section. The red dashed line marks the zero Bernoulli surface, corresponding to the location where the relativistic Bernoulli parameter defined as
\begin{equation}
   Be = -\frac{T^t_{\phantom{t}t} + R^t_{\phantom{t}t} + \rho u^t}{\rho u^t}, 
\end{equation}
vanishes ($Be = 0$). Above it, the gas has sufficient energy to escape to infinity without additional input and within the zero Bernoulli surface, the gas would remain gravitationally bound. Some outflowing material can still contribute to the luminosity of the system \citep[for more details see][]{Kayanikhoo2025}.

The kinetic luminosity of the outflows is computed by integrating outgoing kinetic flux $u^r>0$ above the zero Bernoulli surface, 
\begin{equation}\label{kinenergylum}
    L_{\rm KE} = - 2\pi \int (u_t + \sqrt{-g_{tt}})\rho u^r  \sqrt{-g} d{\theta}\bigg|_{Be>0,\, u^r>0}.
\end{equation}
The kinetic efficiency $L_{\rm KE}/(\mdot c^2)$ with $\mdot$ computed at $20\,r_{\rm g}$ is shown in the top panel of Fig.~\ref{ke_rad_efficiency_mag}. 
For radii $r\geq60\,r_{\rm g}$, the efficiency of the kinetic luminosity increases as the magnetic field strength decreases. Near the star, at $r\leq 20\,r_{\rm g}$, there is no outflow  
and the outgoing kinetic energy remains zero. The outflow power increases as the dipole field weakens.  
In the weak dipole simulation, a thick outflow propagates toward the neutron star's polar axis, forming (collimating) a narrow, optically thin, cone-like region (between the polar axis and the green line shown in Fig.~\ref{outflow}). As the dipole strength increases, this region expands (the angle between the green line and the polar axis widens from left to right in Fig.~\ref{outflow}).

We compute the radial optical depth $\tau_{\rm r}$ as
\begin{equation}\label{tau_r}
        \tau_\mathrm{r}(r)= \int_{r}^{r_{\rm out}} \rho \kappa_{\mathrm{es}} \sqrt{g_{\rm rr}} d{{r}},
\end{equation}
with $r_{\rm out}$ the outer boundary of the simulation and  $\kappa_{\rm es} = 0.34 \, {\rm cm}^2\,{\rm g}^{-1}$ the electron scattering opacity for solar composition. The photosphere with $\tau_{\rm r} = 1$ is shown with the green solid line in each of the panels in Fig.~\ref{outflow}.

We calculate the radiation luminosity $L_{\rm r}$ by integrating the radiation flux over the optically thin region where $\tau_{\rm r} < 1$, allowing radiation to propagate outwards toward the observer (neglecting cosmological effects),

\begin{equation}
    L_{\rm r} =  2 \pi \int_{0}^{\theta} F^r_{\rm rad} \sqrt{-g} d{\theta},
\end{equation}
with $\theta$ located where $\tau_{\rm r} = 1$ (the green line in each frame in Fig.~\ref{outflow}).

In the bottom panel of Fig.~\ref{ke_rad_efficiency_mag}, we observe that $L_{\rm r}/(\mdot c^2)$ for the weak dipole ($10^{10}\,$G) has the highest value and decreases with increasing dipole strength. The radiation increases beyond radius of about $200\,r_{\rm g}$, most likely due to the curvature of the photosphere at larger radii \citep[see Figure 3 in][]{Kayanikhoo2025}, at which more radiation flux flows to the optically thin region. 
 
We estimate the apparent (isotropic) luminosity $L_{\rm iso}$ as
\begin{equation}\label{Eqapparent}
    L_{\rm iso} (\theta) = 4\pi d^2 F^r_{\rm rad}(d), 
\end{equation}
where $d$ is the distance between the observer and the source and $F^r_{\rm rad}(d)$ is radiation flux in the optically thin region at distance $d$. 
Neglecting attenuation, $L_{\rm iso}$ should not depend on the distance once it exceeds a certain critical value, which we take to be $d = 500\,r_{\rm g}$ in this study.

The isotropic luminosity in the polar coordinates, $L_{\rm iso}$, is shown in Fig.~\ref{apparent_mag}. 
The viewing angle at which observable luminosity drops to zero varies with magnetic dipole strength, as a result of a change in the width of the optically thin region and the power of the outflow.

In Figure~\ref{apparent_mag}, we also show that the apparent luminosity peaks along the polar axis and decreases at larger viewing angles. The maximum apparent luminosity in the weak magnetic dipole case ($10^{10}\,$G) reaches approximately $115\,L_{\rm Edd}$, and this peak value diminishes with increasing magnetic field strength, dropping to around $40\,L_{\rm Edd}$ in the strong dipole case ($10^{11}\,$G).

\subsection{Study of accretion rates}
 
In Fig.~\ref{outflowmdot} we show the inflow–outflow structure. 
The outflow increases with increasing accretion rate in each magnetic dipole strength  
both below and above the zero Bernoulli surface.  
The outflow at a high accretion rate propagates to the high latitudes and creates a narrow cone-like, optically thin region ($\tau_{\rm r}<1$) close to the polar axis (between the green line and the polar axis).

The kinetic luminosity efficiency $L_{\rm KE}/(\mdot c^2)$ with $L_{\rm KE}$ computed using Eq.~\ref{kinenergylum}, is plotted in the top panel of Fig.~\ref{ke_rad_efficiency_mdot}. 
The kinetic efficiency increases with the accretion rate and the weakening of the magnetic dipole field. 

The radiation efficiency, defined as $L_{\rm r}/(\dot{M} c^2)$, where $L_{\rm r}$ is computed within the optically thin region characterized by $\tau_{\rm r} < 1$, is presented in the bottom panel of Fig.~\ref{ke_rad_efficiency_mdot}. 
The higher accretion rates correspond to lower radiation efficiencies. At larger accretion rates, significant outflows develop, and as photons cross these dense outflows, they interact with the outflowing material, transferring momentum. This interaction diminishes the radiation efficiency while simultaneously enhancing the kinetic efficiency

The apparent luminosity, calculated using Eq.~\ref{Eqapparent}, is shown in Fig.~\ref{apparmdot} for a range of accretion rates and two magnetic dipole strengths.
An interesting trend is shown in Figure~\ref{apparmdot}: 
as the accretion rate increases, the apparent luminosity also increases, in contrast to the behavior shown in Fig.~\ref{ke_rad_efficiency_mdot}, where the radiation efficiency decreases with increasing accretion rate. This 
can be attributed to the presence of powerful outflows, which lead to strong collimation of the radiation and consequently higher apparent luminosity along certain viewing angles. 
In regions where the outflow obscures the radiation, the observable radiation flux drops to zero.
These results show that a neutron star with a dipole strength in the order of $10^{10}\,$G accreting above 300~Eddington units may power ULXs.

\section{Discussion and conclusions}\label{Sconc}
%
We conducted ten numerical simulations of the super-Eddington accreting magnetised neutron star with the GRRMHD code, \texttt{Koral}. We investigated magnetic dipole strengths between $10^{10} - 10^{11}\,$G and accretion rates of 100, 300, and 1000~$L_{\rm Edd}/c^2$. 

The key parameters extracted from the time-averaged data of the simulations are shown in Fig.~\ref{conclusions}: the magnetospheric radius $R_{\rm m}$, outflow rate $\mdotout$, luminosity $L$ (computed in the optically thin region) and the beaming factor $b$.
The most prominent observed trend is an inverse correlation between the magnetospheric radius, $R_{\rm m}$ and the remaining three parameters, with an increase in $R_{\rm m}$ consistently corresponding to a decrease in each of the other parameters. This suggests that the geometry of the accretion plays a significant role in the observed radiative properties. 

The strength of the magnetic dipole determines the magnetospheric radius of the accretion disk, so that a stronger dipole field leads to disk truncation at larger radii. Consequently, the outflow rate decreases with increasing magnetic dipole strength. 
In simulations for a strong dipole, accreting material is funneled along reconnected magnetic field lines and reaches the neutron star's surface primarily through the polar regions. In contrast, with a weaker dipole, the magnetic field lines open up, allowing material to propagate freely to higher latitudes. Also, more outflows are generated in weak dipole simulations. In these cases, accretion occurs predominantly through the equatorial region of the star (see Fig.~\ref{magavg-grid}). The increased propagation of material to high latitudes enhances turbulence, which in turn amplifies radiation emission.

The observed trend in the final parameter, the inverse of the beaming factor, $1/b$, where $b = L_{\rm iso}/L_{\rm r}$
is directly linked to the outflow rate.  
A strong magnetic field leads to a reduced outflow and weaker beaming. The $b$ values calculated at two different radii show the same trend, but the value computed at the smaller radius (250~$r_{\rm g}$) is lower than that at the larger radius (500~$r_{\rm g}$).

The results with the different strengths of the magnetic dipole  
show that with weak dipole simulation ($10^{10}\,$G), 
the beaming factor is about 0.017, which 
corresponds to the apparent luminosity about 120~$L_{\rm Edd}$ computed at the radius 500~$r_{\rm g}$, with 
values are gradually changing with the magnetic dipole strength. For the dipole one order of magnitude stronger ($10^{11}\,$G), the beaming factor increases by almost an order of magnitude to approximately 0.08 
and the corresponding apparent luminosity about 40~$L_{\rm Edd}$. 

The accretion rate significantly influences the radius at which the accretion disk is truncated. The radius at which the ratio of ram pressure to magnetic pressure equals unity decreases with increasing accretion rate.  
This variation in $R_{\rm m}$ is more pronounced in the simulation with the magnetic dipole of $3\times10^{10}\,$G. In simulations with a weak magnetic dipole, where $R_{\rm m}$ is already very small at low accretion rate, an increase in the accretion rate does not significantly impact the magnetosphere.

An inverse relationship is also found between the magnetospheric radius and the outflow rate. A smaller magnetospheric radius corresponds to a higher outflow rate. Consequently, the largest outflow rate in our simulations is achieved at accretion rate of approximately 1000 Eddington luminosity units ($\mdot_{1000}$).

The final parameter, $b$, follows directly from the outflow. The structure of the accretion disk and column, along with the propagation of the outflow, leads to the formation of a cone-like optically thin region near the polar axis, which determines how the radiation is beamed. In the high accretion rate simulation with $\mdot = 1000\, L_{\rm Edd}/c^2$, the beaming factor is approximately 0.008 in the weak magnetic dipole simulation ($10^{10}\,$G), and about 0.01 in the strong magnetic dipole simulation ($3\times10^{10}\,$G). These beaming factors correspond to apparent luminosities of 185 and 220~$L_{\rm Edd}$, respectively, consistent with the luminosity of ULXs.

Our simulations suggest that the ULXs are most likely neutron stars  with a moderate magnetic field, on the order of $10^{10}\,$G, accreting at super-Eddington rates. The accretion rate strongly influences the dynamics of outflows, beaming. Consequently the apparent luminosity increases with the accretion rate. 
The lowest accretion rate in our simulations, about $100\,L_{\rm Edd}/c^2$, probably falls short of ULX requirements even in the presence of the most favorable magnetic dipole field (as low as $10^{10}\,$G).

We implemented an energy-reflecting surface with an albedo of 0.75, meaning that all the energy reaching the neutron star surface is converted into radiation energy, of which 75\% is reflected. However, the boundary condition at the neutron star surface does not function as intended.
With high accretion rates and strong magnetic fields, 
much of  the reflected radiation is advected by the flow and carried below the inner boundary, thus reducing the radiative efficiency by up to two orders of magnitude. We conjecture that with a realistic implementation of the albedo the radiative luminosity close to the axis would increase relative to the already high values reported here.

Another caveat is that, although the simulations are run for a sufficient duration to allow the system to reach a stable state, and time-averaged data is collected over an extended period, in setups with a strong magnetic field it would be beneficial to conduct longer simulations as accretion tends to stabilize at later times in these cases.

 In this study, we investigate the impacts of two key parameters, magnetic dipole strength and accretion rate, with the aim of identifying the combinations that yield luminosities consistent with those observed in ULXs. To draw more definitive conclusions regarding the relationship between the luminosity and the accretion structure, simulations with a wider variation of magnetic dipole strengths and accretion rates are required. Additionally, several other parameters warrant further investigation, including the multipole magnetic field, the compactness of the central object, and the magnetic field structure of the torus.

\begin{acknowledgments}
Research in CAMK was supported in part by the Polish National Center for Science grant 2019/33/B/ST9/01564. F.K. acknowledges the Polish National Center for Science grant no. 2023/49/N/ST9/01398 and the Operational Programme Just Transformation - Vouchers for Universities - Moravian-Silesian region CZ.10.03.01/00/23-042/0000390. M.Č. acknowledges the Czech Science Foundation (GAČR) grant No. 21-06825X and the support by the International Space Science Institute (ISSI) in Bern, which hosted the International Team project No. 495 (Feeding the spinning top) with its inspiring discussions. 
We gratefully acknowledge Polish high-performance computing infrastructure PLGrid (HPC Center: ACK Cyfronet AGH) for providing computer facilities and support within computational grant no. PLG/2023/016648.
\end{acknowledgments}

\software{Koral, Numpy \citep{numpy2011, numpy2020}, Scipy \citep{scipy2022}, Matplotlib \citep{matplotlib2007CSE}, and Pandas \citep{mckinney2010data}.}

\bibliography{nsulxs}{}
\bibliographystyle{aasjournal}

\end{document}